%% file: main.tex
\documentclass[apj,numberedappendix,twocolumn,twocolappendix,floatfix]{openjournal}

\DeclareRobustCommand{\VAN}[3]{#2}
\let\VANthebibliography\thebibliography
\def\thebibliography{\DeclareRobustCommand{\VAN}[3]{##3}\VANthebibliography}

\usepackage{amsmath}
\usepackage{amssymb}
\usepackage{booktabs}
\usepackage{lipsum}
\usepackage[T1]{fontenc}
\usepackage{graphicx}
\usepackage[breaklinks,colorlinks,citecolor=blue,urlcolor=blue,linkcolor=blue]{hyperref}
\usepackage{cleveref}
\usepackage{multirow}
\usepackage{newtxtext,newtxmath}
\usepackage{orcidlink}
\usepackage{pifont}
\usepackage{natbib}
\usepackage{xparse}
\usepackage[dvipsnames]{xcolor}

\usepackage{physics}
\usepackage{enumitem}

\newcommand\vell[0]{\boldsymbol{\ell}}

\defcitealias{Bonnerot+2022}{BPL22}
\defcitealias{Tomida+2013}{T+13}

\begin{document}

\title{Resolving the (Debate about) Nozzle Shocks in Tidal Disruption Events}

\author{\vspace{-1.3cm}
Zachary L. Andalman\,\orcidlink{0000-0001-5064-1269}$^{1,\star}$,
Eliot Quataert\,\orcidlink{0000-0001-9185-5044}$^{1}$,
Eric R. Coughlin\,\orcidlink{0000-0003-3765-6401}$^{2}$,
and C. J. Nixon\,\orcidlink{0000-0002-2137-4146}$^{3}$\\
}

\affiliation{
$^{1}$Department of Astrophysical Sciences, Princeton University, 4 Ivy Lane, 08540, Princeton, NJ, USA\\
$^{2}$Department of Physics, Syracuse University, Crouse Drive, 13210, Syracuse, NY, USA\\
$^{3}$School of Physics and Astronomy, University of Leeds, Sir William Henry Bragg Building, Woodhouse Ln, Leeds, LS2 9JT, United Kingdom
}
\thanks{$^{\star}$E-mail:zack.andalman@princeton.edu}

\begin{abstract}
\input{sections/00_abstract}
\end{abstract}

\maketitle

\section{Introduction}
\input{sections/01_introduction}

\section{Methods}
\input{sections/02_methods}
\label{sec:methods}

\section{Results}
\input{sections/03_results}
\label{sec:results}

\section{Discussion}
\input{sections/04_discussion}
\label{sec:discussion}

\section{Conclusion}
\input{sections/05_conclusion}
\label{sec:conclusions}

\section*{Acknowledgements}
\input{sections/06_acknowledgements}

\section*{Data Availability}
\input{sections/07_data}

\bibliographystyle{aasjournal}
\bibliography{main}

\appendix
\input{sections/08_appendix}

\end{document}

%% file: sections/00_abstract.tex
When a star passes within the Roche limit of a supermassive black hole (SMBH), it is pulled apart by the BH's tidal field in a tidal disruption event (TDE). The resulting flare is powered by the circularization and accretion of bound stellar debris, which initially returns to the BH on eccentric orbits in a thin debris stream. The returning fluid elements follow inclined orbits that converge near pericenter, resulting in extreme vertical compression to scales $10^{-4}~R_\odot$ and the formation of a nozzle shock. Dissipation at the nozzle shock may affect circularization by altering the properties of the debris stream, but its role is the subject of ongoing debate. We develop an idealized model for the debris stream evolution combining 3D smoothed-particle hydrodynamics simulations, the semi-analytic affine model, and 1D finite-volume hydrodynamic simulations. Because our model is computationally cheap, we can unambiguously resolve the nozzle shock, use a realistic equation of state, and follow the debris stream evolution at many different times. Near peak fallback, Hydrogen recombination and molecular Hydrogen formation broaden the stream by a factor $\sim 5$, enhancing dissipation at the nozzle. However, the dissipation is still insufficient to directly circularize the debris by in-plane pressure gradients. Instead, the thicker stream substantially increases the likelihood that the stream self-intersects on the second orbit, despite relativistic nodal precession. The stream properties at self-intersection are sensitive to dissipation at the nozzle and the timing of focal points where the ballistic trajectories of the debris converge. Our results clarify the nozzle shock's role in circularization in TDEs, providing a foundation for more realistic circularization and emission models.

%% file: sections/01_introduction.tex
Stars in galactic nuclei occasionally fall onto nearly-parabolic orbits that bring them within the Roche limit of the central supermassive black hole (SMBH), where they are pulled apart by the BH's tidal field in a tidal disruption event (TDE) \citep{Hills1975, Frank&Rees1976, Lightman&Shapiro1977, Rees1988, Magorrian&Tremaine1999, Gezari2021, Mockler+2025}. A portion of the stellar material is bound to the BH and falls back in a thin debris stream on timescales of weeks to years \citep{Lacy+1982, Evans&Kochanek1989}. This fallback powers bright flares in the X-ray \citep{Bade+1996, Komossa&Bade1999, Saxton+2012}, UV \citep{Gezari+2006, Gezari+2008} and optical \citep{vanVelzen+2011, Arcavi+2014, Holoien+2014}, with bolometric peak luminosities around $10^{44}{\rm erg/s}$ and a light curve that often follows the theoretical rate of mass return to the BH $\propto t^{-5/3}$ \citep{Phinney1989}. The severity of the disruption is characterized by the penetration factor $\beta \equiv r_{\rm t} / r_{\rm p}$, where $r_{\rm p}$ is the pericenter distance of the orbit, $r_{\rm t} = R_* Q^{1/3}$ is the tidal radius, and $Q \equiv M_\bullet / M_*$ is the BH-to-star mass ratio.

Although TDEs are rare, with a rate about $10^{-4}/{\rm gal}/{\rm yr}$ \citep{Rees1988, Magorrian&Tremaine1999, Wang&Merritt2004, Stone&Metzger2016, Gezari2021}, there are over 100 candidates \citep[see the \textit{Open TDE Catalog};][]{Auchettl+2017}. The number of observed events is expected to grow substantially with new data from the Legacy Survey of Space and Time/Rubin Observatory \citep[LSST;][]{Ivezic+2019, Bricman&Gomboc2020}. These data will make TDEs a powerful tool to address broader astrophysical questions. TDEs can constrain the properties of otherwise quiescent supermassive BHs \citep{Rees1988, Komossa2015, Mockler+2019}, advancing the study of BH fueling modes, growth mechanisms, and evolution. TDE rates and host galaxy properties are sensitive to the kinematics of galactic nuclei \citep{Frank&Rees1976, Lightman&Shapiro1977, LawSmith+2017, Pfister+2020}. Finally, TDEs are accretion physics laboratories where we can observe the same BH in different feeding regimes from super- to sub-Eddington \citep{Evans&Kochanek1989, Strubbe&Quataert2009, Dai+2018}.

These applications require a detailed understanding of how properties of the disruption and fallback map onto the observed flare. Optical/UV emission is typically interpreted as intrinsic X-ray emission reprocessed by optically-thick outflows \citep{Roth+2016, Dai+2018}. The intrinsic emission is produced as the stellar debris circularizes \citep{Piran+2015} and then accretes onto the BH, although only a fraction of the bound debris may ultimately be accreted \citep{Ayal+2000, Andalman+2022, Ryu+2023}. 

\citet{Rees1988} originally proposed that the debris is circularized by the stream-stream collisions. As the debris stream falls back towards the BH, it is deflected onto a self-intersecting trajectory by relativistic apsidal precession. At the self-intersection, the radial velocities of the incoming and outgoing streams in the rest frame of the BH are equal and opposite with speeds $\sim 0.01c$ \citep{Lu&Bonnerot2020} for a $\beta=1$ disruption by a $10^6~M_\odot$ BH. If all the energy at the shock could be radiated instantaneously, it would produce a luminosity $10^{43}~{\rm erg/s}$ for a typical fallback rate $\dot{M}_{\rm fb} \sim M_\odot/{\rm yr}$.

In practice, the radiative efficiency of the self-intersection is limited by several factors. First, the incoming and outgoing debris streams may have different widths, so only a fraction of the fallback material participates in the self-intersection \citep{Kochanek1994}. Second, if the BH has angular momentum, then the stream is deflected onto a different orbital plane by relativistic nodal precession, leading to a vertical offset at self-intersection or even a complete miss \citep{Kochanek1994, Dai+2013, Hayasaki+2016, Jankovic+2024}, i.e. a ``dark year'' for TDEs \citep{Guillochon&RamirezRuiz2015}, delaying the self-intersection for several orbits \citep{Stone&Loeb2012}. 

Third, the gas is optically thick, so a majority of the radiation energy produced at the shock goes into adiabatic expansion, as seen in local hydrodynamic simulations of the self-intersection \citep{Lee+1996, Kim+1999, Jiang+2016, Lu&Bonnerot2020, Huang+2023, Jankovic+2024}. An alternative source of luminosity is secondary circularization and accretion shocks in the self-intersection outflow, for instance between fluid elements launched on prograde and retrograde orbits with respect to the BH \citep{Bonnerot&Lu2020, Bonnerot+2021}.

The self-intersection may not occur continuously throughout the duration of the fallback. If the fallback is disrupted by the self-intersection, then the self-intersection will be modulated on a timescale of order the Keplerian period at the self-intersection radius \citep{Sadowski+2016, Curd+2021, Andalman+2022}. Once a substantial disk forms, the debris stream can be mixed into the disk by fluid instabilities before it reaches the self-intersection radius \citep{Andalman+2022, Steinberg&Stone2024}, although the efficiency of this mechanism depends on the detailed properties of the debris stream.

Another potentially important effect for circularization occurs at pericenter, prior to self-intersection, where the ballistic focusing of the orbits acts as an effective nozzle \citep{Evans&Kochanek1989, Kochanek1994}. \citet{Evans&Kochanek1989} showed that the flow converges supersonically, and speculated that the nozzle might contribute to circularization through the formation of oblique shocks, altering the orbital velocities. This process is analogous to ballistic focusing in deep initial disruptions $\beta \gg 1$ \citep{Carter&Luminet1982, Luminet&Marck1985, Brassart&Luminet2008, Stone+2013}. In this sense the returning debris can be thought of as a disruption with effective penetration factor $\beta \gtrsim 100$ \citep{Kubli+2025}. However, even if the nozzle shock does not directly circularize the debris, it may affect circularization by altering the properties of the debris stream and therefore the stream-stream and stream-disk interactions.

Over the past few decades, 3D global simulations have found that the debris stream broadens significantly after the nozzle \citep{Lee&Kim1996, Ayal+2000, Guillochon+2014, Shiokawa+2015, Ryu+2023, Steinberg&Stone2024}. The broadening is attributed to a combination of differential apsidal precession across the width of the stream \citep{Evans&Kochanek1989, Gafton&Rosswog2019} and dissipation at the nozzle shock. However, the latter may be dominated by numerical dissipation, because the post-nozzle stream width is not converged at the resolution of most 3D simulations \citep{Price+2024, Huang+2024}.

\citet{Bonnerot&Lu2022} analyzed the nozzle shock in detail using 2D hydrodynamic simulations of a cross section of the debris stream in the orbital frame. They found that the nozzle shock does not produce significant net expansion of the debris stream. Their result was corroborated by recent 3D smoothed-particle hydrodynamics (SPH) simulations with $\gtrsim 10^{10}$ particles \citep{Kubli+2025} \citep[see also][for an alternative approach using adaptive particle refinement]{FitzHu+2025}. However, these works do not consider the effects of Hydrogen (H) recombination and molecular Hydrogen (${\rm H_2}$) formation in the debris stream \citep{Kochanek1994}, which increases the energy available to be dissipated at the nozzle \citep{Coughlin2023, Steinberg&Stone2024}.

In this paper, we develop an idealized model for the debris stream evolution combining 3D SPH simulations, the semi-analytic affine model, and 1D finite-volume hydrodynamic simulations. We draw heavily from previous semi-analytical work \citep{Kochanek1994, Coughlin+2016a, Coughlin+2016b, Bonnerot+2022, Coughlin2023}. Because our model is computationally cheap, we can unambiguously resolve the nozzle shock, use a realistic equation of state, and follow the debris stream evolution at many different times.

In the remainder of this paper, we first (Section~2) describe each step of our model and justify our assumptions. In Section~3, we present the results of our model. In Section~4, we discuss the implications of our results for debris circularization and the limitations of our work. We conclude in Section~5.

%% file: sections/02_methods.tex
\subsection{Overview of the model}

In the initial disruption, the gas is dense and ionized, so the stellar debris can be modeled accurately with a $\gamma$-law EOS. For this step we use the 3D smoothed-particle hydrodynamics (SPH) code {\sc phantom} \citep{Price+2018}. Accurately capturing the dynamics of the debris' return to pericenter requires extreme numbers of particles \citep{Kubli+2025}, but accurate results for the disruption of the star and the formation of a debris stream require at most a few million particles, at least for low-$\beta$ encounters as we consider here \citep{Norman+2021}.

As the debris stream expands and cools, recombination and ${\rm H_2}$ formation become important. Rather than run multiple SPH simulations with different EOSs, we follow the debris stream using the semi-analytic affine model developed by \citet{Kochanek1994} and refined by \citet{Bonnerot+2022}. The affine model breaks down near pericenter due to the formation of a nozzle shock, which violates the model's assumption of homologous evolution. Just before pericenter, we transition to 1D hydrodynamics simulations in \textsc{Athena++} \citep{Stone+2020}. Near the nozzle, the evolution in the orbital plane is dominated by the BH gravitational field. We approximate the in-plane evolution as ballistic, which makes it trivial to couple to our 1D simulations. The gravitational field does not necessarily dominate the in-plane evolution after the nozzle, so we map the results back into the affine model to evaluate the stream conditions at self-intersection. 

\begin{figure*}
    \centering
    \includegraphics[width=0.8\linewidth]{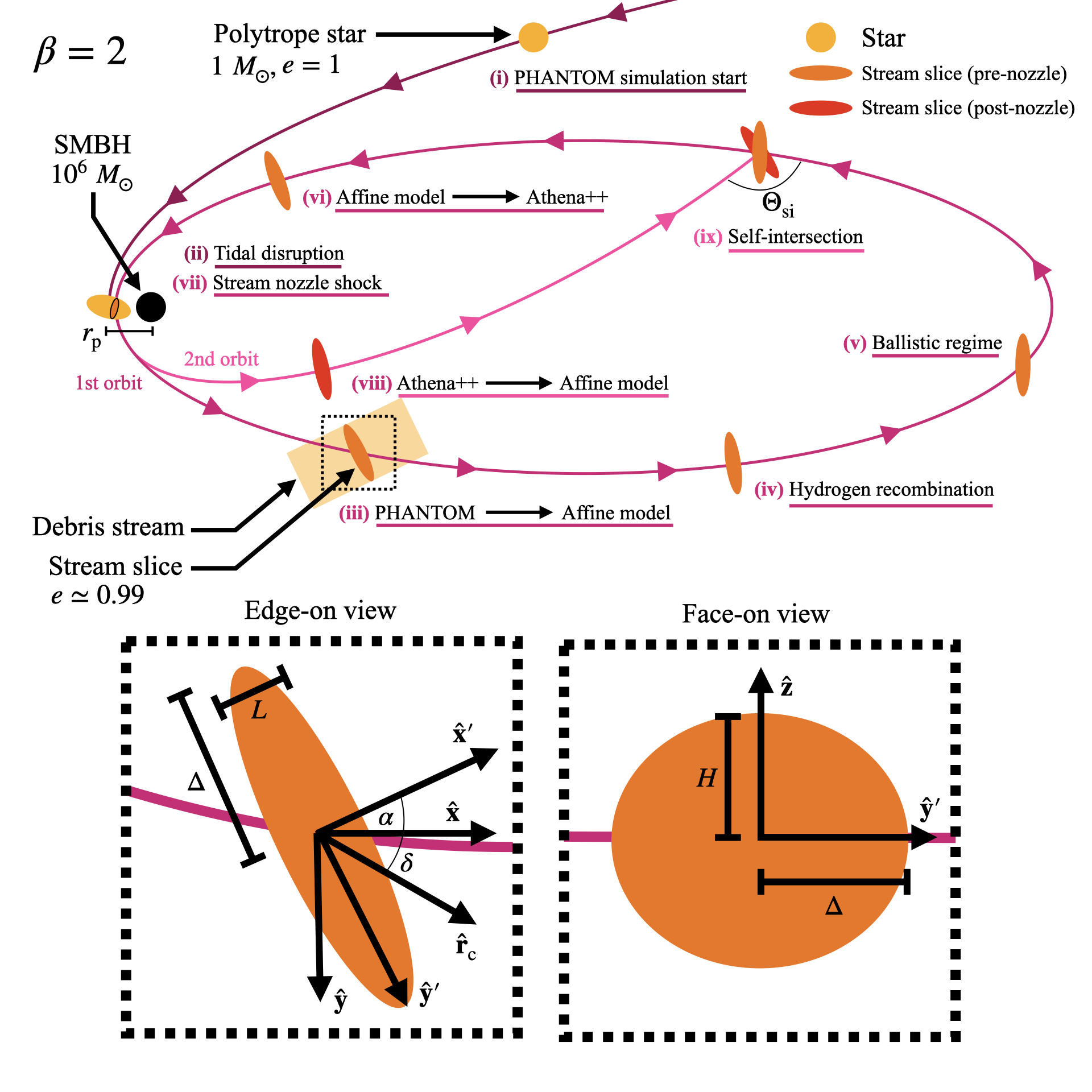}
    \caption{A schematic diagram of the TDE problem, highlighting key steps in the evolution of the stellar debris and our modeling approach: (i) we model the initial $\beta=2$ disruption of a $1~M_\odot$ star by a $10^6~M_\odot$ SMBH using the 3D SPH code {\sc phantom}; (ii) the star gets tidally disrupted near pericenter; (iii) once the stellar debris forms a stream, we use a snapshot of the {\sc phantom} simulation to initialize the affine model for the evolution of a stream slice; (iv) the release of latent heat associated with Hydrogen recombination significantly expands the stream slice; (v) as the stream slice approaches apocenter, it enters the ballistic regime where its dynamics are dominated by the tidal field of the SMBH; (vi) once the stream slice falls back to $10~r_{\rm p}$, we use the affine model to initialize a 1D \textsc{Athena++} simulation; (vii) during the second pericenter passage, the stream slice undergoes a nozzle shock and the bulk of the relativistic apsidal precession occurs; (viii) when the stream returns to $10~r_{\rm p}$, we map the \textsc{Athena++} simulation back into the affine model; (ix) due to the previous apsidal precession, outgoing and incoming stream slices collide at an angle $\Theta_{\rm si}$, dissipating orbital energy and driving circularization. In the bottom of the diagram, we show our coordinate system with origin $\vb{r}_{\rm c}$ and stream slice orthogonal to $\vu{x}'$. On the left side, we show an edge-on view of the stream slice from the $+\vu{z}$-direction. On the right side, we show a face-on view of the stream slice from the $+\vu{x}'$-direction. The unit vector $\vu{x}'$ forms angles $\alpha$ and $\delta$ relative to $\vu{x}$ and $\vu{r}_{\rm c}$ respectively. $H$, $\Delta$, and $L$ describe the dimensions of the stream slice in the $\vu{x}'$, $\vu{y}'$, and $\vu{z}$ directions respectively.}
    \label{fig:diagram}
\end{figure*}

In Figure~\ref{fig:diagram}, we show a schematic diagram of the TDE problem, highlighting key steps in the evolution of the stellar debris and our modeling approach. In Sections~\ref{sec:init_dis}, \ref{sec:affine}, and \ref{sec:athena}, we describe our methods for each phase of the evolution. In Section~\ref{sec:eos}, we explain our implementation of a realistic EOS.

\subsection{The initial disruption in {\sc phantom}}
\label{sec:init_dis}

We simulate the initial disruption of a solar-mass star by a $10^6~M_\odot$ BH in the SPH code {\sc phantom} \citep{Price+2018}, which has been widely used in numerical studies of TDEs \citep[e.g.][]{Coughlin&Nixon2015, Bonnerot+2016, Coughlin+2016a, Bonnerot&Lu2020, Miles+2020, Norman+2021, Andalman+2022, Cufari+2022, Jankovic&Gomboc2023, Jankovic+2024}. For shallow TDEs $\beta \lesssim 4$, 3D SPH simulations accurately capture the formation of the debris stream and the spread in orbital energy, which is underestimated by the standard frozen-in approximation \citep[e.g.][]{Norman+2021}.

We model the SMBH as a Newtonian gravitational potential. We include self-gravity using a k-D tree algorithm, computing self-gravitational forces directly when the opening angle is $\ge 0.5$ \citep[][see \citealt{Price+2018} for details]{Barnes&Hut1986}. We use the standard artificial viscosity switch in {\sc phantom} \citep[an adapted version of the switch developed by][see \citealt{Price+2018} for details]{Cullen&Dehnen2010} with the switch applied to both the $\alpha^{\rm AV}$ and $\beta^{\rm AV}$ terms as suggested by \cite{Chen&Nixon2025}. We employ a $\gamma$-law EOS with pressure related to density by $p = K \rho^{\gamma}$ with adiabatic exponent $\gamma = 5/3$ appropriate for an ideal monoatomic gas.

We take the star to have a density profile of that of an $n=3$ polytrope. The $n=3$ polytrope is a good approximation for the density profile of a partially radiative solar mass star \citep{Golightly+2019}, following the Eddington standard model \citep{Hansen&Kawaler1994}. Many previous works use $n=1.5$ polytropes \citep{Nolthenius&Katz1982, Bicknell&Gingold1983, Evans&Kochanek1989, Guillochon+2014, Bonnerot&Lu2022, Kubli+2025}, which are good approximations for the density profile of fully convective low mass stars $\lesssim 0.5~M_\odot$. We also analyze an $n=1.5$ polytrope star and find no significant change in the qualitative results, so we focus our attention on the $n=3$ polytrope case.

For an $n=3$ polytrope with a $\gamma = 5/3$ equation of state, the entropy function $K$ increases with distance from the center of the star to ensure hydrostatic equilibrium. During the {\sc phantom} simulation, we hold the value of $K$ fixed for each particle, since we do not expect significant shocks during the initial disruption \citep{Coughlin&Nixon2022}, so any non-adiabatic heating would be largely a numerical rather than a physical effect.

To initialize the star, we place $16 \times 10^6$ particles with a uniform distribution in spherical polar angles (equal probability in azimuth and cosine of the zenith angle) and a radial distribution that matches the radial density profile of the polytrope. This configuration of particles is relaxed in isolation for 10 dynamical times ($\sqrt{R_\star^3/GM_\star}$) of the star with a damping force. By this time the total kinetic energy has reduced to just below $10^{-7}$ of the total thermal energy of the star. 

The relaxed star is then placed in the gravitational potential of the SMBH. The star is positioned with its center of mass at a distance of $5~r_{\rm t}$ from the SMBH. We chose the velocity such that the center-of-mass is on a parabolic orbit with pericenter distance $r_{\rm p} = r_{\rm t} / \beta$, where $\beta$ is the penetration factor. We use a penetration factor $\beta = 2$, because $n=3$ polytropes are only partially disrupted for $\beta = 1$ \citep{Guillochon&RamirezRuiz2013}. 

We take a snapshot of the simulation after $2.7~{\rm hr}$ when the star reaches a distance $10~r_{\rm p}$ from the SMBH after the disruption. We record the position $\vb{r}$, velocity $\vb{v}$, density $\rho$, pressure $p$, and smoothing length $\ell$ of each SPH particle. From the position and velocity, we compute the specific orbital energy and angular momentum. Bound particles $\varepsilon < 0$ fall back to the SMBH after one orbital period, $t_{\rm fb} = 2\pi G M_\bullet (-2 \varepsilon)^{-3/2}$. We use these data to initialize the affine model described in the following subsection.

\subsection{The affine model}
\label{sec:affine}

The affine model is an expansion of the fluid equations about the center of each stream slice, assuming homologous evolution. The model is affine because the transformation between the initial and final coordinates of any fluid element is linear in orbital frame. In the affine model, we describe a stream slice by its thickness $L$, in-plane width $\Delta$, and vertical width $H$ (Fig.~\ref{fig:diagram}), so the stream slice density scales as $\rho \propto (L \Delta H)^{-1}$. The gas obeys an EOS defined by adiabatic indices $\Gamma_1 \equiv \partial \ln p / \partial \ln \rho |_s$ and $\Gamma_3 \equiv 1 + \partial \ln T / \partial \ln \rho |_s$, which we use to evolve the entropy proxy $K \equiv p / \rho^{5/3}$ and the temperature respectively. We denote dimensionless quantities with a hat, defined by dividing each variable by its initial value. We work in a rotating coordinate system $\vb{r}'$ where $\vu{x}'$ (resp. $\vu{y}'$) is oriented parallel (resp. perpendicular) to the long axis of the debris stream and evolves due to tidal torques.

We briefly highlight how our version of the affine model differs from \citet{Kochanek1994} and \citet{Bonnerot+2022}, leaving the more complete description to Appendix~\ref{sec:affine_eqs}. In comparison to \citet{Kochanek1994}: we use a Newtonian rather than a pseudo-Newtonian potential for the BH; we evolve the entropy proxy rather than the specific internal energy; and we don't include viscous heating. In comparison to both works, we assume a stream density profile obtained by linearly scaling a hydrostatic cylinder. This gives the numerical prefactor in the pressure gradient force, similar to \citet{Coughlin2023}. We also set up our model to interface with an arbitrary EOS. The affine model equations become stiff at the nozzle due to extreme vertical compression, so we use an implicit 5th-order Runge-Kutta method of the Radau IIA family as implemented by \texttt{scipy.integrate.solve\_ivp}.

We use snapshots of our {\sc phantom} simulations at $2.7~{\rm hr}$ to initialize the affine model for each stream slice. We discard particles with large smoothing lengths $> 0.003~r_{\rm p}$ which contribute most to the numerical error (noise) in the subsequent binning procedure, roughly $1\%$ of the $16\times 10^6$ particles. We bin the remaining particles by their specific orbital energy. From each bin, we select the $64$ highest-density particles to represent the center of the stream slice and compute their mean position $\vb{r}_{\rm c}$, velocity $\vb{v}_{\rm c}$, density $\rho_{\rm c}$, and pressure $p_{\rm c}$. The number $64$ is similar to the typical number of neighbor particles that contribute to the default smoothing kernel in {\sc phantom} \citep[58 for the cubic spline kernel][]{Price+2018}. We fix $z_{\rm c} = v_{z, {\rm c}} = 0$, since we expect the center-of-mass orbits to lie in the $xy$-plane by symmetry. 

We combine the data from all bins to construct arrays that describe fluid quantities along the long axis of the debris stream. We smooth the resulting arrays with a $100$ element Gaussian to reduce noise. In the smoothed arrays, the radial coordinate increases monotonically with specific orbital energy. We compute $\vell = -\partial \vb{r}_{\rm c} / \partial \varepsilon$ and $\partial_t \vell = -\partial \vb{v}_{\rm c} / \partial \varepsilon$ using first-order finite differences and construct the primed coordinates $\vu{x}' = \vell / \ell$ and $\vu{y}' = \vu{z} \cross \vu{x}'$, from which we compute the affine model orbital parameters $\lambda$, $\Omega$, and $\alpha$ (Fig.~\ref{fig:diagram}). At each coordinate along the stream, the stream slice plane is defined by $\cos \alpha (x - x_{\rm c}) + \sin \alpha (y - y_{\rm c}) = 0$. We select the $2^{14} = 16384$ particles closest to the slice plane as members of the stream slice.

As mentioned earlier (Sec.~\ref{sec:init_dis}), the initial entropy in the star is stratified as $K \equiv p/\rho^{5/3} \propto \rho^{-1/3}$. However, within each stream slice, $K$ is maximized at the center, so deviations from the central value are second order in cylindrical radius. Therefore, we can reasonably approximate the initial entropy in each stream slice as a constant \citep{Coughlin2023}. This implies that the stream slice density profiles satisfy the cylindrical Lane-Emden equation (Eq.~\ref{eq:lane_emden}) with Emden index $n=1.5=1/(5/3-1)$, assuming that stream slices reach cylindrical hydrostatic equilibrium after the disruption. As a \textit{post hoc} justification, such profiles closely match the stream slice density profiles in the simulation, especially at small cylindrical radii.

Our affine model equations are dimensionless, so the absolute scale of the stream slice dimensions is a degree of freedom. We chose the initial $\hat{L} = 1$, which means that pressure gradients and self-gravity are balanced for $\hat{H} = \hat{\Delta} = 1$ (Eqs.~\ref{eq:H} and \ref{eq:Delta}). We determine the dimensionless stream widths by fitting a cylindrical polytrope density profile to the stream slice particles. The profile is scaled by $\hat{H}$ and $\hat{\Delta}$ relative to the equilibrium width $\alpha_{\rm le} \xi_{\rm max}$. 

We minimize an $\ell_2$-norm loss function using the Nelder-Mead method as implemented in \texttt{scipy.optimize.minimize} \citep{Gao&Lixing2010}. We allow an offset in the $y'$-direction, which is typically small relative to $\Delta$. We determine $\partial_t \hat{H}$, $\partial_t \hat{\Delta}$, and $\hat{v}_\parallel$ using the maximum likelihood slope of the appropriate velocity and position components for particles in the stream slice. At the end of this procedure, we have a full set of initial conditions for the affine model.

\subsection{The returning debris in \textsc{Athena++}}
\label{sec:athena}

As the bound stellar debris returns to pericenter, it undergoes extreme vertical compression and a nozzle shock. Shock formation cannot be captured by the affine model because it breaks the assumption of homologous evolution. We model this phase of the evolution using 1D simulations in the hydrodynamics code \textsc{Athena++}, representing the central column of each stream slice. These simulations are coupled to the affine model describing the in-plane evolution.

\textsc{Athena++} evolves fluid variables on a discrete mesh according to the Euler equations. We use \textsc{Athena++}'s default MUSCL-Hancock scheme, a second-order Godunov method, with the Harten-Lax-van-Leer-Contact (HLLC) Riemann solver \citep{Harten+1983}. The problem is symmetric about the orbital plane, so we impose a reflective inner boundary at $z=0$. For the outer boundary at $z = 0.5~r_{\rm p} = 25~R_\odot$, we impose a Neumann boundary condition. We use $2^{14} = 16384$ logarithmically spaced cells with a minimum cell size $\Delta z \simeq 1.2\times 10^{-8}~r_{\rm p} \simeq 7.0 \times 10^{-7}~R_\odot$ at the center of the stream slice and a ratio $1.00062$ between neighboring cells.

In one dimension, the Euler equations are,
\begin{equation}
	\partial_t \mathcal{U} + \partial_z \mathcal{F} = \mathcal{S}
\end{equation}
where the conserved variables are $\mathcal{U} \equiv [\rho, \rho v, (1/2) v^2 + e]$, the fluxes are $\mathcal{F} \equiv [\rho v, \rho v^2 + p, (1/2) \rho v^3 + v (e + p)]$, and $e$ is the internal energy density. The source terms $\mathcal{S} \equiv [\mathcal{S}_{\rho}, \mathcal{S}_{\rm mom}, \mathcal{S}_{\rm kin} + \mathcal{S}_{\rm thm}]$ include contributions from the tidal acceleration $a_{\rm t} = -GM_\bullet z / r_{\rm c}^3$ and the in-plane evolution, parameterized by the dimensionless in-plane area $\hat{A} \equiv \hat{L} \hat{\Delta}$,
\begin{align}
    \mathcal{S}_\rho =\ & -\rho \partial_t \ln \hat{A}\\
    \mathcal{S}_{\rm mom} =\ & \rho a_{\rm t} - \rho v \partial_t \ln \hat{A}\\
    \mathcal{S}_{\rm kin} =\ & \rho v a_{\rm t} - (1/2) \rho v^2 \partial_t \ln \hat{A}\\
    \mathcal{S}_{\rm thm} =\ & -(1 + p / e) e \partial_t \ln \hat{A}
\end{align}
The factor $1 + p/e$ in the thermal energy source term ensures that the in-plane evolution is adiabatic, since $1 + p/e = \partial \ln e / \partial \ln \rho |_s$ from the first law of thermodynamics. We don't include a source term contribution from self-gravitational acceleration because it is negligible relative to the other source terms at the start time of the \textsc{Athena++} simulations.

The \textsc{Athena++} simulations run from when the stream slice falls back to $10~r_{\rm p}$ in the affine model to when it returns to $10~r_{\rm p}$ after the nozzle. We use the affine model to precompute $r_{\rm c}$, $\hat{A}$, and $\partial_t \hat{A}$. During the simulation, we linearly interpolate from these arrays to calculate the source terms. For each stream slice, we set up an $n=1.5$ cylindrical polytrope density and pressure profile using $H$ and $K$ from the affine model. We set the initial velocity in each cell according to homologous collapse $v_z = (\partial_t \hat{H} / \hat{H}) z$.

We set density and pressure floors at $10^{-16} M_\bullet / r_{\rm p}^{3} \simeq 4.7\times 10^{-15} {\rm g/cm^3}$ and $10^{-22} G M_\bullet^2 / r_{\rm p}^4 \simeq 0.18~{\rm erg/cm^3}$ respectively. We track the floor material using a passive scalar mask which is equal to unity in the physical region and exponentially falls to zero in the floor region. We use this mask to avoid applying source terms to floor material. This avoids issues at the outer boundary where tidal acceleration would otherwise create an evacuated region.

We use the conditions at the end of the \textsc{Athena++} simulations at $2.7~{\rm hr}$ to reinitialize the affine model. Initializing too soon $\lesssim 1~{\rm hr}$ after the nozzle produces inaccurate results because the evolution in \textsc{Athena++} immediately after the nozzle is far from homologous; in particular, we tend to underestimate the central pressure at late times compared to the \textsc{Athena++} simulation. However, initializing too late $\gtrsim 1~{\rm day}$ might miss substantial deviations from ballistic in-plane evolution.

We refer to the affine models initialized from the {\sc phantom} and \textsc{Athena++} snapshots as the first and second affine models respectively. We compute $\hat{H}$ from the central density via $\hat{H} = (\rho_{\rm 0, c} / \rho_{\rm c}) \hat{A}$ and use numerical differentiation to compute the time derivative. We use the central density and pressure to compute the entropy proxy and temperature. We take the values of other quantities from the first affine model. 

After the nozzle shock, the stream profile may no longer be described by a rescaled cylindrical polytrope. Therefore, the pressure gradient term in the second affine model may be off by an order-unity factor related to the exact post-shock profile. However, for an idealized model, the second affine model provides a reasonable qualitative description of the post-nozzle evolution.

\subsection{Equations of state}
\label{sec:eos}

Most numerical studies of TDEs assume a $\gamma$-law EOS $p \propto \rho^\gamma$ with adiabatic exponent $\gamma = 5/3$ appropriate for an ideal monoatomic gas. However, changes in ionization state and chemical composition modify the thermodynamics of the debris stream. Partial ionization states are associated with low adiabatic indices $\Gamma \approx 1 + 2 k_{\rm B} T / \chi$ \citep{Cox1963, Kasen&RamirezRuiz2010}, where $\chi$ is the ionization energy, because $p\dd V$ work goes into changing the ionization state rather than changing the temperature and pressure. Similar reasoning applies to transitional states in a chemical reaction.

Our formulation of the affine model supports arbitrary EOSs given expressions for the effective adiabatic indices as a function of density and temperature. We use two additional EOSs. Our first EOS accounts for H recombination and ionization using the Saha equation. Our second EOS additionally accounts for Helium (He) recombination and ionization, ${\rm H_2}$ formation and dissociation, and radiation pressure using the partition function from \citet{Tomida+2013} (hereafter \citetalias{Tomida+2013}). We implement the second EOS in \texttt{Athena++} using the built-in general EOS module \citep{Coleman2020}. 

As the stellar debris leaves pericenter, it expands and cools adiabatically, eventually reaching temperatures $\sim 10^4~{\rm K}$ where H recombination occurs \citep{Kochanek1994, Kasen&RamirezRuiz2010, Coughlin2023}. The recombination energy ($\chi_{\rm H} = 13.6~{\rm eV}$ per Hydrogen atom) can do $p\dd V$ work to expand the stream, increasing the energy available to dissipate at the nozzle shock \citep{Steinberg&Stone2024}. From the first law of thermodynamics and the Saha equation, the adiabatic indices are,
\begin{align}
    \Gamma_{1, {\rm Saha}} =\ & 1 + \frac{1 + 2 (3/2 + u_{\rm H}) f(x_{\rm i}) - (3/2) f(x_{\rm i})}{(3/2) + (3/2 + u_{\rm H})^2 f(x_{\rm i})}
    \label{eq:gam1}\\
    \Gamma_{3, {\rm Saha}} =\ & 1 + \frac{1 + (3/2 + u_{\rm H}) f(x_{\rm i})}{(3/2) + (3/2 + u_{\rm H})^2 f(x_{\rm i})}
    \label{eq:gam3}
\end{align}
where $x_{\rm i} \equiv n_{\rm e} / n_{\rm H}$ is the ionization fraction, $u_{\rm H} \equiv \chi_{\rm H} / k_{\rm B} T$, and $f(x_{\rm i}) \equiv x_{\rm i} (1 - x_{\rm i}) (2 - x_{\rm i})^{-1} (1 + x_{\rm i})^{-1}$. Our Equation~\ref{eq:gam3} is equivalent to Equation~16 of \citet{Kasen&RamirezRuiz2010} when He ionization and radiation are ignored.

If there is a significant He fraction, then He will also recombine at higher temperatures \citep{Kochanek1994, Kasen&RamirezRuiz2010}. We assume a He fraction $Y=0.3$ close to the solar value. At lower temperatures, we expect ${\rm H_2}$ to form via the three-body reaction $3{\rm H} \to {\rm H_2} + {\rm H}$. The reaction occurs quickly relative to the local dynamical time \citep{Kochanek1994, Forrey2013}, so we assume ${\rm H_2}$ is in statistical equilibrium with the other species. Later, we show that ${\rm H_2}$ formation maintains a temperature $\gtrsim 3000~{\rm K}$ in the debris stream (Fig.~\ref{fig:phasespace}), preventing the formation of dust that would otherwise dramatically enhance the opacity \citep{Kochanek1994}. Radiation pressure should also be accounted for at the nozzle shock, where it becomes significant relative to gas pressure.

\begin{figure*}
    \centering
    \includegraphics[width=0.67\linewidth]{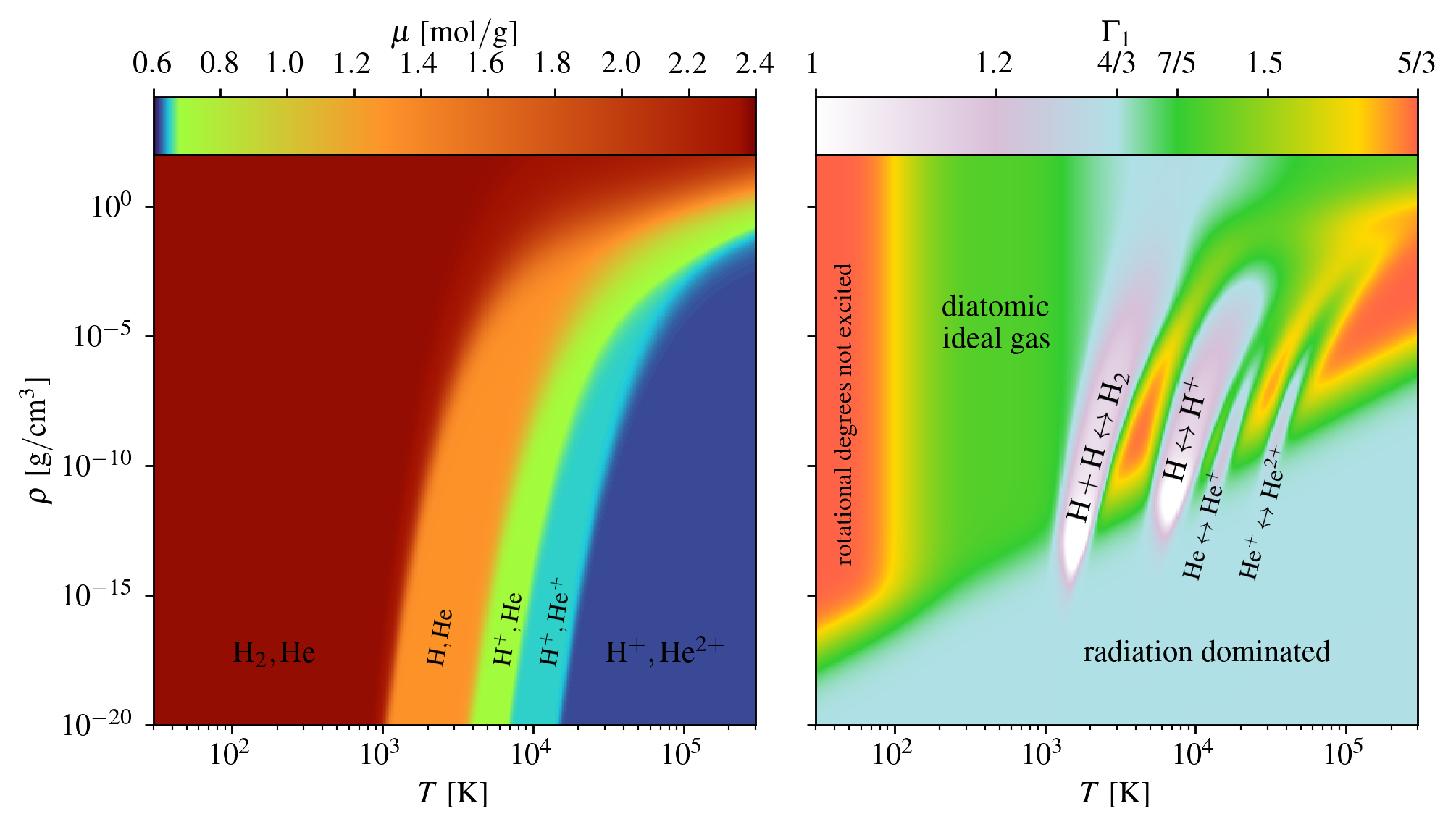}
    \caption{Contour plots of mean molecular weight (left panel) and effective adiabatic index $\Gamma_1 \equiv \partial \ln p / \partial \ln \rho |_s$ (right panel) as a function of density and temperature using the \citetalias{Tomida+2013} EOS. In the left panel, we label different regions based on the dominant chemical species: (i), ${\rm H}^+$ and ${\rm He}^{2+}$ (dark blue, $\mu \simeq 0.62$); (ii), ${\rm H}^+$ and ${\rm He}^{+}$ (light blue, $\mu \simeq 0.65$); (iii), ${\rm H}^+$ and ${\rm He}$ (green, $\mu \simeq 0.68$); (iv), ${\rm H}$ and ${\rm He}$ (orange, $\mu \simeq 1.29$); and (v), ${\rm H_2}$ and ${\rm He}$ (red, $\mu \simeq 2.35$). In regions of partial ionization or dissociation, the effective adiabatic index approaches unity because $p\dd V$ work goes into the chemical reaction rather than changing the temperature and pressure.}
    \label{fig:tomida}
\end{figure*}

Including these effects makes the EOS expensive to compute on-the-fly. Instead, we use the tabulated EOS of \citetalias{Tomida+2013}, originally developed for simulations of protostellar cores. Later works used the same EOS in simulations of binary stars \citep{Pejcha+2016} and TDEs \citep{Steinberg&Stone2024}. In their Appendix~1, \citetalias{Tomida+2013} provide the partition function that includes the relevant chemical processes. At high densities $\rho \gtrsim 0.1{\rm g/cm^3}$, this treatment of the EOS becomes inaccurate due to interactions between particles, quantum effects, pressure ionization of Hydrogen, and contributions from other chemical species. However, even at the nozzle shock, the stream slice density never rises significantly above this threshold. We modify the partition function to include the contributions to the pressure, internal energy, and entropy from radiation similar to \citet{Pejcha+2016}. In Figure~\ref{fig:tomida}, we show our tables for mean molecular weight and adiabatic index $\Gamma_1$.

In their Appendix~1, \citetalias{Tomida+2013} describe the procedure for computing thermodynamic variables from the partition function. We use this procedure to generate tables of the first and second derivatives of the partition function over a grid of densities and temperatures. We compute second derivatives of the partition function by numerical differentiation rather than root-finding. All other thermodynamic variables can be derived from the partition function derivatives. The code to generate our tables is available online.\footnote{\href{https://zandalman.com/publications/Andalman+2026a/}{zandalman.com/publications/Andalman+2026a/}.}

We use \textsc{Athena++}'s built-in general EOS module \citep{Coleman2020} to run simulations with the \citetalias{Tomida+2013} EOS. We invert our EOS table numerically to generate the required inverted tables: $\log (p/\varepsilon)[\log \rho, \log \varepsilon]$, $\log (\varepsilon/p)[\log \rho, \log (p / \rho)]$, and $\log \Gamma_1 [\log \rho, \log (p / \rho)]$. 

At low temperatures $\lesssim 85~{\rm K}$, the internal energy density asymptotes to the zero-point energy (ZPE) associated with the vibrational mode of ${\rm H_2}$. These temperatures are never reached by the debris stream, but they are occasionally reached in the floor material. This makes it difficult to invert the EOS because the temperature is a weak function of the internal energy density. As a workaround, we subtract the ZPE from the internal energy density in the EOS table.

%% file: sections/03_results.tex
\subsection{Initial disruption results}
\label{sec:init_dis_res}

We use snapshots of our {\sc phantom} simulations at $2.7~{\rm hr}$ to initialize our affine model calculations. In the first couple days after the disruption, pressure gradient and self-gravitational forces along the direction of tidal elongation still have some effect \citep{Kochanek1994}. Ideally, we would initialize the affine model using a snapshot $\gtrsim 50~{\rm hr}$, when we expect the specific orbital energy distribution to fully converge \citep{Fancher+2023}. However, the later snapshots occur after significant H recombination, especially for the most bound stream slices. By initializing the affine model with the snapshot at $2.7~{\rm hr}$, we capture all of the H recombination and accept small errors in the final orbits of the debris.

\begin{figure*}
    \centering
    \includegraphics[width=0.8\linewidth]{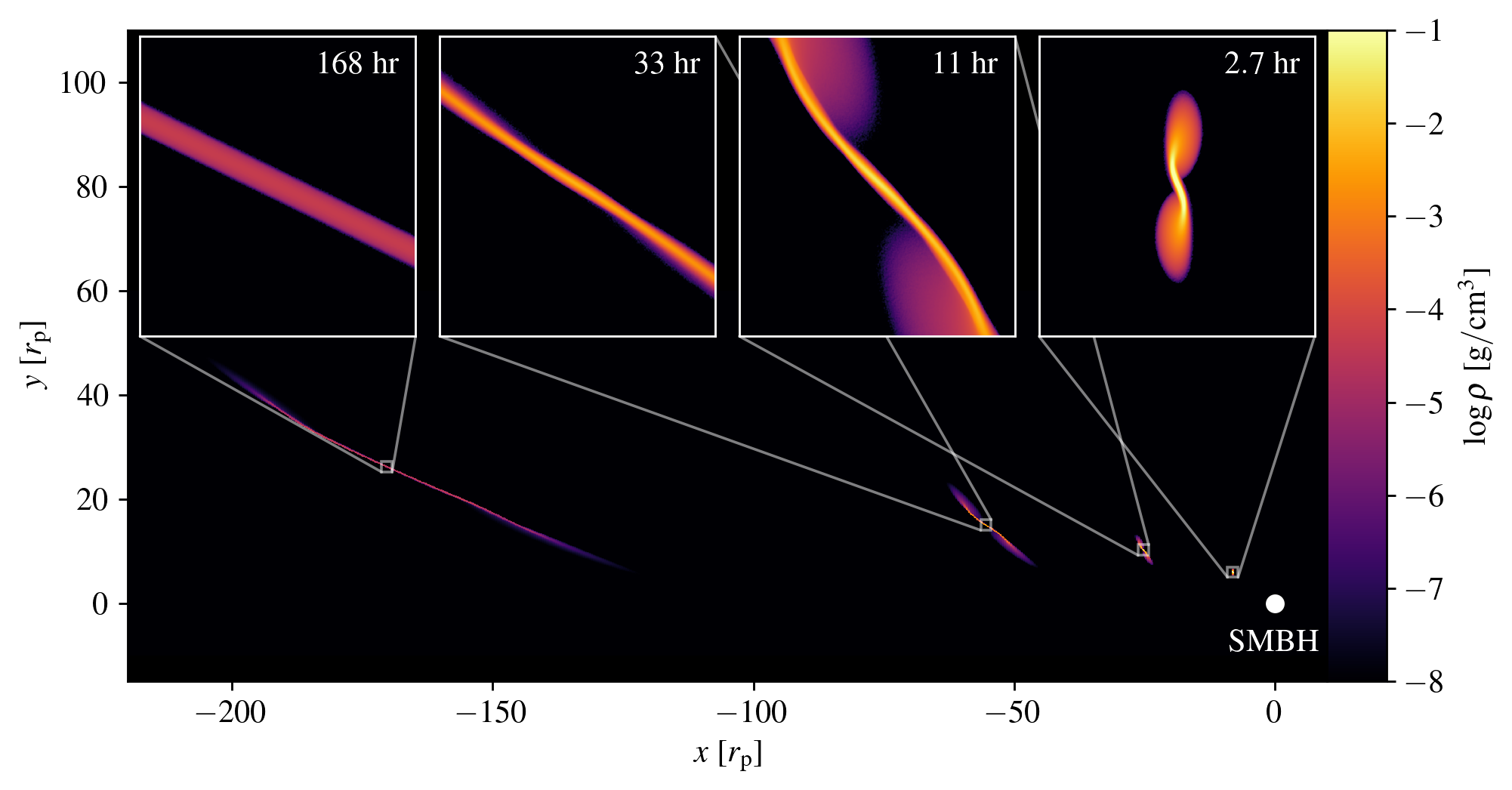}
    \caption{The logarithmic density in the $xy$-plane averaged over the $z$-direction and weighted by mass from our simulation of the initial disruption in {\sc phantom}. In the main panel, we superimpose snapshots at 2.7, 11, 33, and 168 hours. In the inset panels, we show a zoom-in of each snapshot centered on the marginally bound debris. Each inset panel has a side length $2~r_{\rm p}$. We indicate the SMBH using a white dot and an annotation. After the initial disruption, the tidal field elongates the stellar debris into a quasi-cylindrical structure which is well-described by the affine model. We use the snapshot at $2.7~{\rm hr}$ to initialize our affine model calculations.  }
    \label{fig:sph}
\end{figure*}

In Figure~\ref{fig:sph}, we show the density in the $xy$-plane in 4 snapshots at 2.7, 11, 33, and 168 hours. As a test of our procedure, we initialize the same stream slice with different snapshots and evolve them using the affine model with a $\gamma$-law EOS (not shown). The evolution of orbital parameters $\lambda$, $\Omega$, and $\alpha$ is independent of the snapshot used in the initialization. However, the stream width evolution is sensitive to the initial snapshot because the amplitude of quadrupolar oscillations (Sec.~\ref{sec:first_orb}) evolves over the {\sc phantom} simulation due to interference with other oscillation modes not captured by the affine model (Nixon et al. in prep.). We leave the detailed characterization of stream oscillations to future work, acknowledging that there is some uncertainty in our stream widths associated with our choice of initial snapshot.

\begin{figure}
    \centering
    \includegraphics[width=\linewidth]{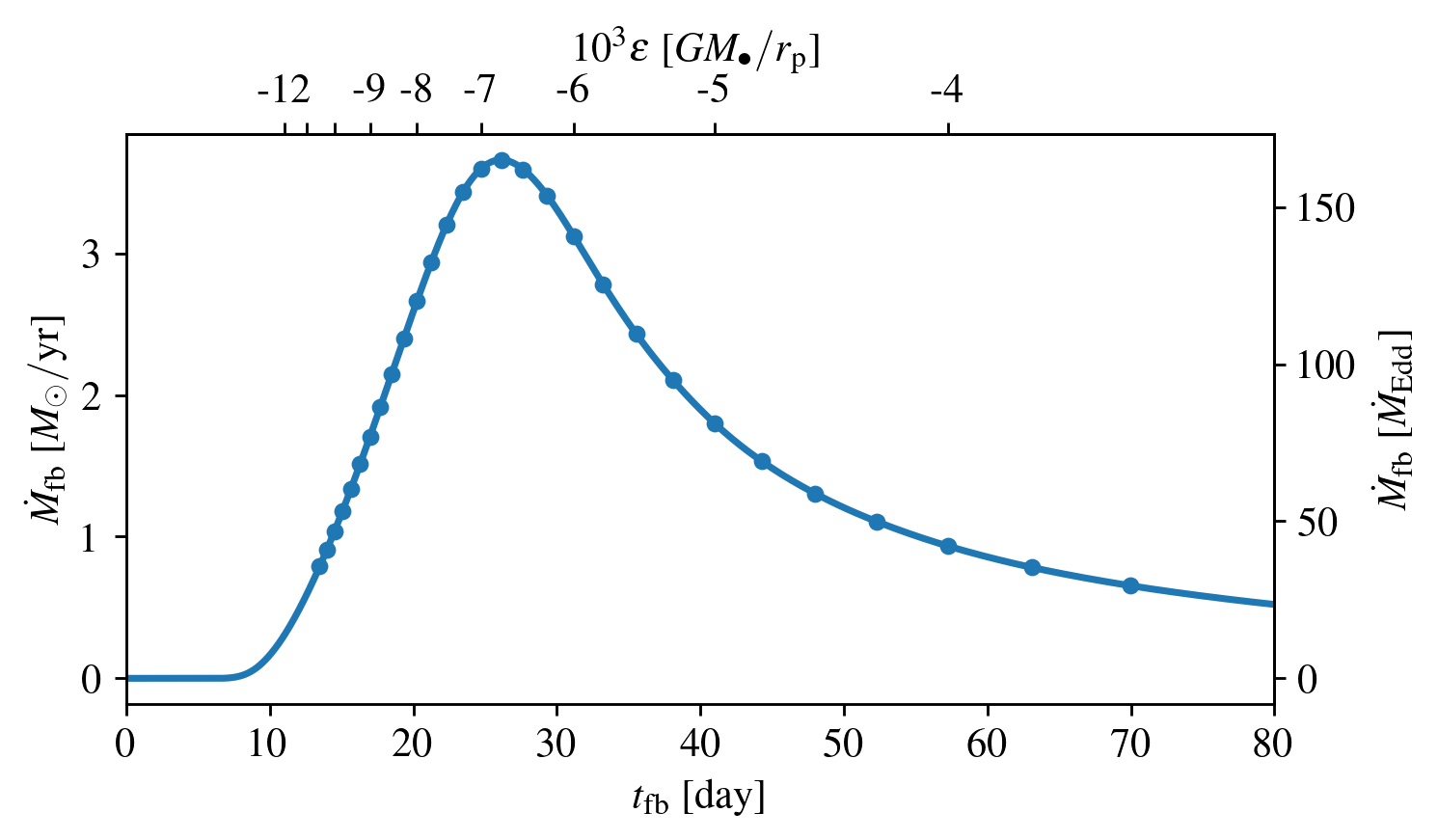}
    \caption{The mass fallback rate as a function of time (bottom $x$-axis) and specific orbital energy (top $x$-axis) in units of $M_\odot/{\rm yr}$ (left $y$-axis) and Eddington accretion rate (right $y$-axis) from our {\sc phantom} simulation. Each dot corresponds to a stream slice we simulate using the affine model and \textsc{Athena++}. The peak fallback rate $160~\dot{M}_{\rm Edd}$ occurs $27~{\rm day}$ after the star passes pericenter, corresponding to orbital energy $-0.0067~G M_\bullet / r_{\rm p}$. }
    \label{fig:mdot}
\end{figure}

In Figure~\ref{fig:mdot}, we show the mass fallback rate as a function of time and specific orbital energy. We express the fallback rate in physical units and units of the Eddington accretion rate given by
\begin{equation}
	\dot{M}_{\rm Edd} = \frac{L_{\rm Edd}}{\epsilon_{\rm r} c^2} \simeq 0.02 M_6~M_\odot/{\rm yr}, \quad L_{\rm Edd} = \frac{4\pi G M_\bullet m_{\rm p} c}{\sigma_{\rm T}}
\end{equation}
where $\epsilon_{\rm r} \simeq 0.1$ is a fiducial radiative efficiency, $m_{\rm p}$ is the proton mass, $\sigma_{\rm T} \simeq 6.65 \times 10^{-25}~{\rm cm^2}$ is the Thomson scattering cross section, and $M_6 \equiv M_\bullet / 10^6 M_\odot$. The peak fallback rate $160~\dot{M}_{\rm Edd}$ occurs $27~{\rm days}$ after the pericenter passage, corresponding to orbital energy $-0.0067~G M_\bullet / r_{\rm p}$. 

We initialize 29 stream slices at linearly spaced specific orbital energies from $-0.0105$ to $-0.0035 GM_\bullet/r_{\rm p}$, represented by dots in Figure~\ref{fig:mdot}. We treat the stream slice with specific orbital energy $-0.00675~G M_\bullet / r_{\rm p}$ as a fiducial case, because it lies close to the peak fallback specific orbital energy.

In the initial snapshot, the stream maintains a constant width of about a solar radius, except near the bound tail, where it broadens to a few solar radii. Most stream slice profiles are close to an $n=1.5$ cylindrical polytrope with the same central density and pressure, giving initial $\hat{H} \simeq \hat{\Delta} \simeq 1$. However, the most bound stream slices $\varepsilon \le -0.008 G M_\bullet / r_{\rm p}$ have larger initial in-plane widths by up to a factor of a few. These stream slices have lower densities and enter the ballistic regime at smaller radii, so they are more susceptible to tidal stretching in the early post-disruption phase, when the stream slice normals are nearly perpendicular to the radial direction.

\subsection{Affine model evolution on the first orbit}
\label{sec:first_orb}

\begin{figure}
    \centering
    \includegraphics[width=\linewidth]{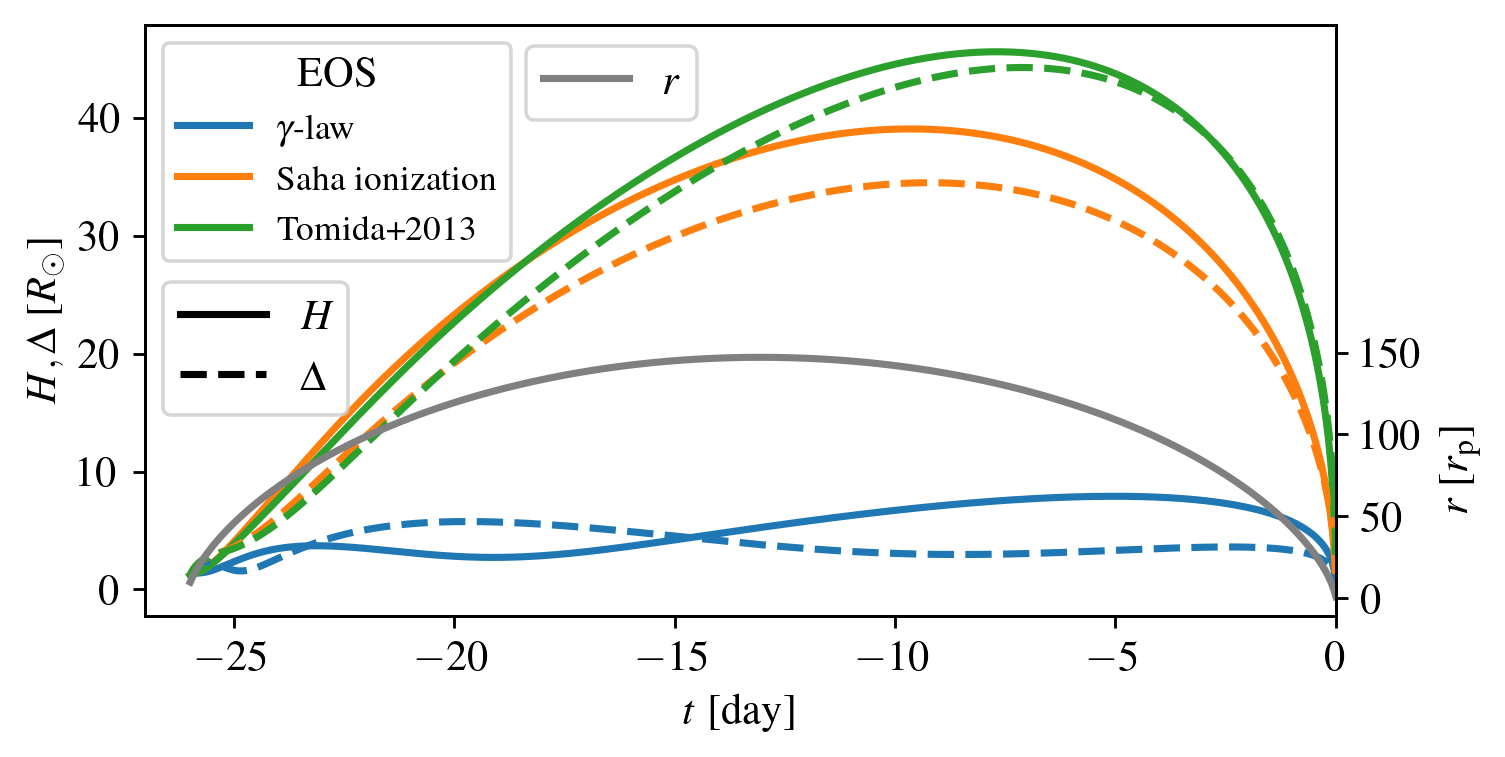}
    \caption{The vertical ($H$; solid lines) and in-plane ($\Delta$; dashed lines) stream widths as a function of time for the fiducial stream slice on the first orbit. We show results with $\gamma$-law (blue lines), Saha ionization (orange lines), and \citetalias{Tomida+2013} EOSs (green lines). We also show the orbital radius (grey line). Chemical processes augment pressure gradients and promote stream expansion by keeping the debris stream hot as it expands. Including H recombination increases the maximum pre-nozzle stream widths by a factor $\sim 5$. Including other chemical processes adds an additional order-unity factor. }
    \label{fig:dim}
\end{figure}

After the initial disruption, we evolve the stream slices using the affine model. In Figure~\ref{fig:dim}, we show the evolution of the stream widths on the first orbit for the fiducial stream slice. Chemical processes augment pressure gradients and promote stream expansion by keeping the debris stream hot as it expands, or equivalently, by lowering the effective adiabatic index (App.~\ref{sec:toy_model}). Including H recombination increases the maximum stream widths by a factor $\sim 5$. Including other chemical processes adds an additional order-unity factor. 

The difference in the stream widths between the H recombination EOS run and the \citetalias{Tomida+2013} EOS run is driven by ${\rm H_2}$ formation rather than ${\rm He}$ recombination. When the stream material reaches the temperatures for ${\rm He}$ recombination, the high gas density shifts the partial ionization region to higher temperatures where the ionization energy is a smaller fraction of the internal energy, suppressing its impact on gas thermodynamics.

The vertical gravitational potential energy at the maximum vertical stream width is equivalent to the kinetic energy available to dissipate at the nozzle, so we expect a much stronger nozzle shock when H recombination is included \citep{Steinberg&Stone2024}. 

With a $\gamma$-law EOS, the stream widths oscillate out-of-phase about their quasi-equilibrium values during the hydrostatic regime. These oscillations correspond to the $m=2$ quadrupolar oscillations identified in previous work on the debris stream \citep{Kochanek1994, Coughlin+2016a, Coughlin+2016b, Bonnerot+2022, Pacuraru+2025}, which are seeded by the initial disruption. The oscillation period is roughly constant in logarithmic time. This is consistent with \citet{Coughlin2023}'s analysis of the $m=0$ breathing mode, which showed that the oscillation is overstable. Other stream slices (not shown) also exhibit oscillations, except for the most bound stream slices, which enter the ballistic regime before the first oscillation. With the other EOSs, the oscillations are overpowered by stream expansion driven by chemical processes. 

The initial disruption can also seed perturbations along the length of the stream, which can lead to the fragmentation of the stream into self-gravitating clumps \citep{Coughlin&Nixon2015}. \citet{Coughlin+2016b} argue that the stream is stable to fragmentation for adiabatic indices $\Gamma_1 < 5/3$, supported by the SPH simulations of \citet{Coughlin+2016a} at different adiabatic indices. This condition is always satisfied with a realistic EOS, so we consider fragmentation unlikely in the bound debris. Fragmentation occurs more readily in the unbound debris once it becomes optically thin enough for significant radiative cooling.

\begin{figure}
    \centering
    \includegraphics[width=\linewidth]{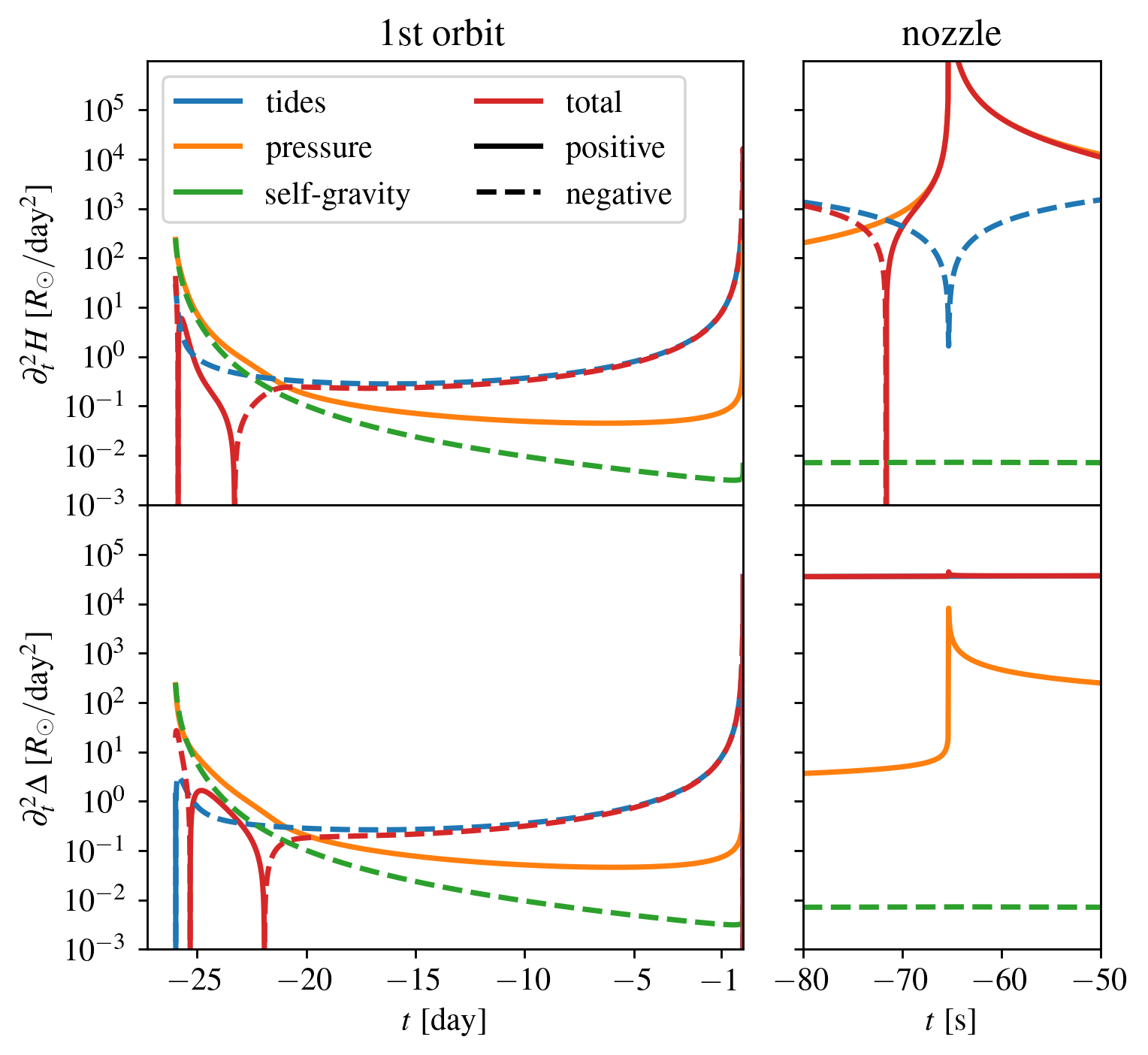}
    \caption{The dimensionless acceleration associated with each term in the affine model equations for $H$ (top panels) and $\Delta$ (bottom panels) as a function of time, with respect to the pericenter passage, on the first orbit for the fiducial stream slice with a \citetalias{Tomida+2013} EOS. The contributions include tides (blue lines), pressure gradients (orange lines), and self-gravity (green lines). We also show their sum (red lines). Positive and negative contributions are indicated by solid and dashed lines respectively. In the right panels, we zoom in on the evolution near the nozzle using data from the \textsc{Athena++} simulations. In the hydrostatic regime, pressure gradient and self-gravitational forces are balanced. In the ballistic regime, the tidal force dominates. The vertical collapse at the nozzle takes place over seconds. }
    \label{fig:evol}
\end{figure}

In Figure~\ref{fig:evol}, we show the dimensionless acceleration associated with each term in affine model Equations~\ref{eq:H} and \ref{eq:Delta} as a function of time for the fiducial stream slice with a \citetalias{Tomida+2013} EOS. In this section, we focus on the left panels, which cover the first orbit of the debris. In the hydrostatic regime, pressure gradient and self-gravitational forces are initially balanced. However, once H recombination begins, pressure gradient forces dominate by a factor of a few. About 5 days after the disruption, the stream slice transitions to the ballistic regime, where the tidal force dominates and the stream width accelerations become negative. After this point, self-gravity never becomes dynamically relevant again.

\begin{figure}
    \centering
    \includegraphics[width=\linewidth]{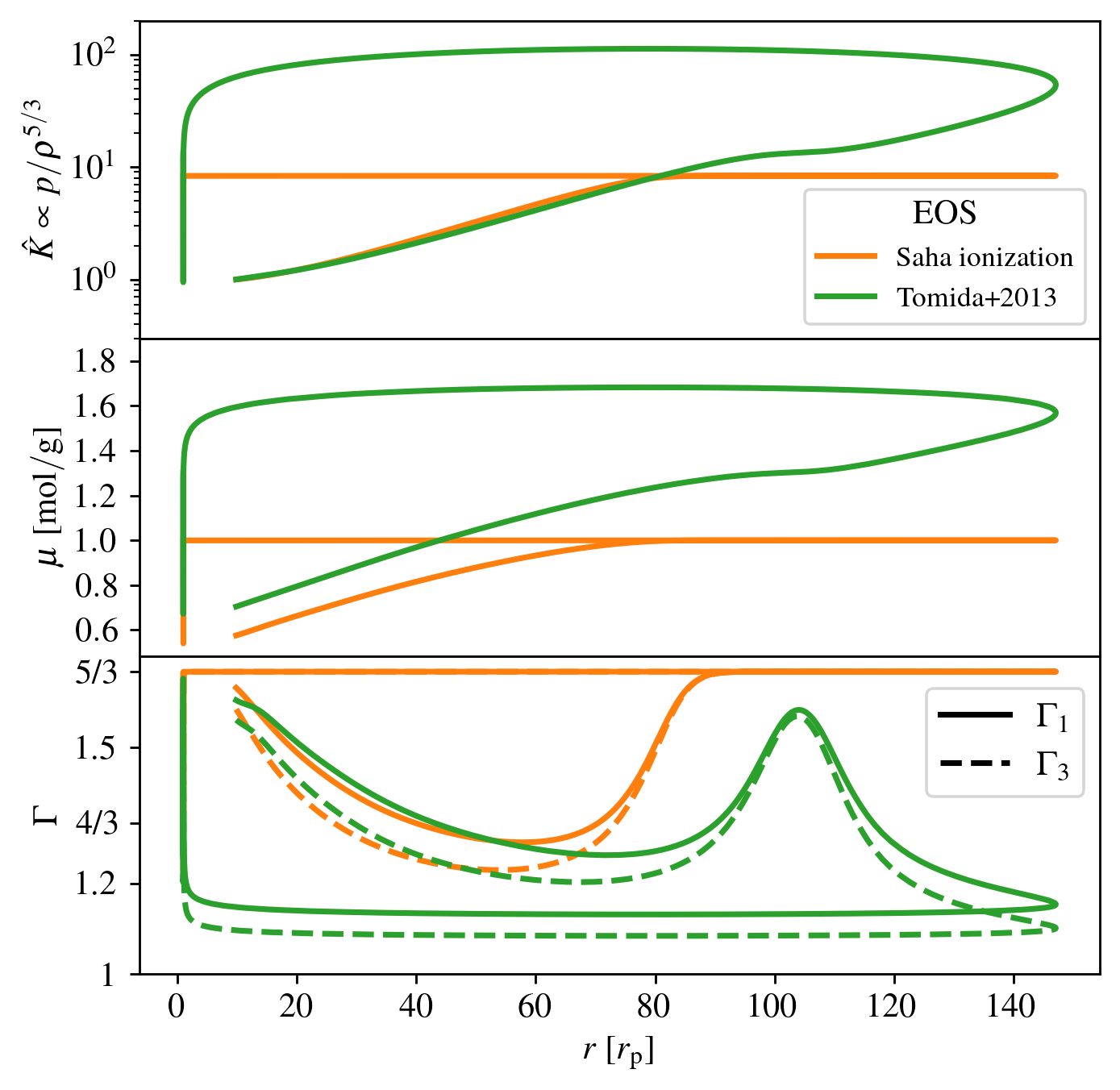}
    \caption{The dimensionless entropy proxy $\hat{K}\propto p / \rho^{5/3}$ (top panel), mean molecular weight (middle panel), and effective adiabatic indices (bottom panel) as a function of radius in the affine model for the fiducial stream slice. In each panel, we indicate the initial conditions (IC) with an annotation. We show results with Saha ionization (orange lines) and \citetalias{Tomida+2013} (green lines) EOSs. For both EOSs, the stream slice completes H recombination near $100~r_{\rm p}$ and remains neutral until the nozzle. For the \citetalias{Tomida+2013} EOS, the stream slice continuously forms or dissociates ${\rm H_2}$. Therefore, ${\rm H_2}$ formation drives stream expansion even after all the Hydrogen has recombined. }
    \label{fig:gam}
\end{figure}

In Figure~\ref{fig:gam}, we show the evolution of the dimensionless entropy proxy, mean molecular weight, and effective adiabatic indices for the fiducial stream slice with Saha ionization and \citetalias{Tomida+2013} EOSs. The stream slice completes H recombination at radii $<100~r_{\rm p}$ and remains neutral until the nozzle. With a \citetalias{Tomida+2013} EOS, the stream slice continuously forms or dissociates ${\rm H_2}$, keeping the effective adiabatic indices low. This lets ${\rm H_2}$ formation drive stream expansion even after all the Hydrogen has recombined. 

The stream widths on the first orbit depend on how much H recombination and ${\rm H_2}$ formation occur during the hydrostatic regime when the stream width is most sensitive to pressure. With a \citetalias{Tomida+2013} EOS, the maximum stream widths on the first orbit increase monotonically with fallback time because less bound stream slices spend more time in the hydrostatic regime. Chemical processes still drive some expansion in the ballistic regime because the ratio of pressure gradient to tidal forces increases as the stream slice recedes to apocenter if $\Gamma_1 \lesssim 1.2$ (App.~\ref{sec:toy_model}).

\begin{figure}
    \centering
    \includegraphics[width=\linewidth]{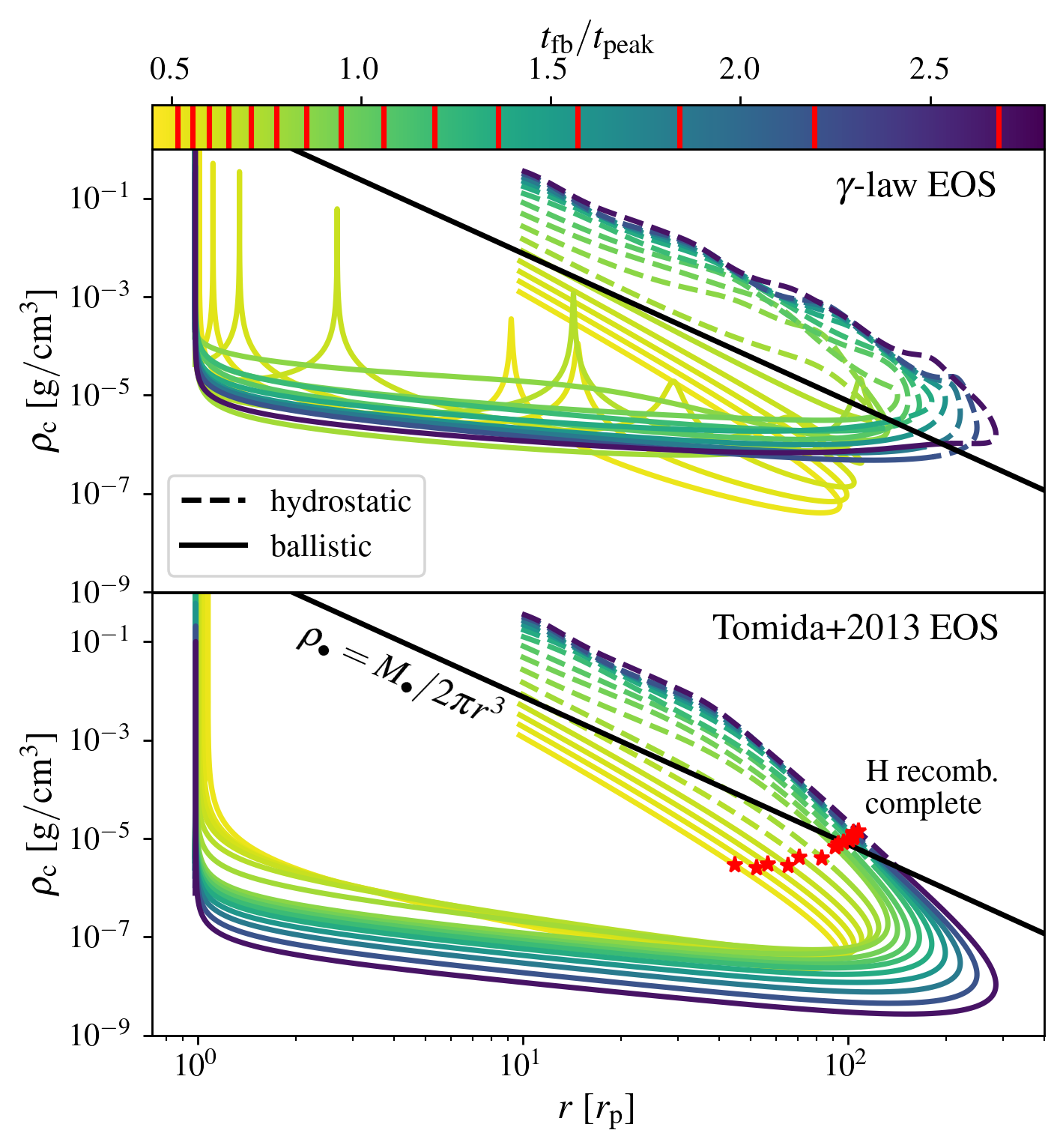}
    \caption{The density as a function of radius in the affine model for 15 stream slices. We indicate the fallback time relative to peak of each stream slice by its color in the color bar. We distinguish between the hydrostatic (dashed lines) and ballistic (solid lines) phases of the evolution. We show results with $\gamma$-law (top panel) and \citetalias{Tomida+2013} (bottom panel) EOSs. For the \citetalias{Tomida+2013} EOS, we indicate the completion of H recombination (red stars). In each panel, we plot the BH density $\rho_\bullet = M_\bullet / (2\pi r^3)$. Most stream slices enter the ballistic regime close to where their density drops below the BH density. The most bound stream slices are already in the ballistic regime in the {\sc phantom} snapshot used to initialize the affine model. For a $\gamma$-law EOS, the most bound stream slices show density peaks at large radii during their return to the BH. These correspond to ballistic focal points, which we define and analyze in Section~\ref{sec:second_orb}.}
    \label{fig:rho}
\end{figure}

In Figure~\ref{fig:rho}, we show the density evolution for 15 stream slices with $\gamma$-law and \citetalias{Tomida+2013} EOSs. We separate the evolution into hydrostatic and ballistic phases by comparing the relevant terms in affine model Equations~\ref{eq:H} and \ref{eq:Delta}. The stream slices enter the ballistic regime close to where their density drops below the BH density $\rho_\bullet = M_\bullet / (2\pi r^3)$ (App.~\ref{sec:toy_model}). In the hydrostatic regime, the stream slice and BH densities have the same scaling with radius for a $\gamma$-law EOS, so they do not enter the ballistic regime until apocenter. With the \citetalias{Tomida+2013} EOS, the rapid expansion triggered by H recombination moves the transition to the ballistic regime before apocenter. 

For a $\gamma$-law EOS, the most bound stream slices in Figure~\ref{fig:rho} show density peaks at large radii during their return to the BH. These peaks correspond to ballistic focal points, which we define and analyze in Section~\ref{sec:second_orb}.

For the \citetalias{Tomida+2013} EOS, we indicate the completion of H recombination with a red star. H recombination completes at later times for less bound stream slices due to their higher initial temperature, but earlier times relative to the transition to the ballistic regime. The least bound stream slices are completely neutral by the time they enter the ballistic regime.

\subsection{Nozzle shock formation and evolution}
\label{sec:nozzle_formation}

When the stream slice reaches $10~r_{\rm p}$ on its way back to the SMBH, we use the affine model to initialize our 1D \textsc{Athena++} simulations. In Figure~\ref{fig:evol}, we zoom in on the evolution near the nozzle. At the nozzle, vertical pressure gradients grow until they reverse the collapse. The reversal takes place over seconds, a small fraction of the orbital period of $26~{\rm days}$. Due to shock dissipation, the pressure after the nozzle is greater than before the nozzle. In the plane, the pressure gradient force increases at the nozzle but never dominates the contribution from tides. This justifies our use of the first affine model to describe the in-plane evolution in the \textsc{Athena++} simulations, which does not capture the effect of dissipation. 

At the nozzle, the vertical stream width is $10^{-4}~R_\odot$ (resp. $2 \times 10^{-4}~R_\odot$) and the in-plane stream width is $3.0~R_\odot$ (resp. $0.6~R_\odot$) with a \citetalias{Tomida+2013} (resp. $\gamma$-law) EOS. This vertical stream width is smaller than the value $10^{-3}~R_\odot$ reported by \citet{Bonnerot&Lu2022} due to stronger tidal forces in our $\beta = 2$ disruption compared to the $\beta = 1$ disruption they consider. 

\begin{figure}
    \centering
    \includegraphics[width=\linewidth]{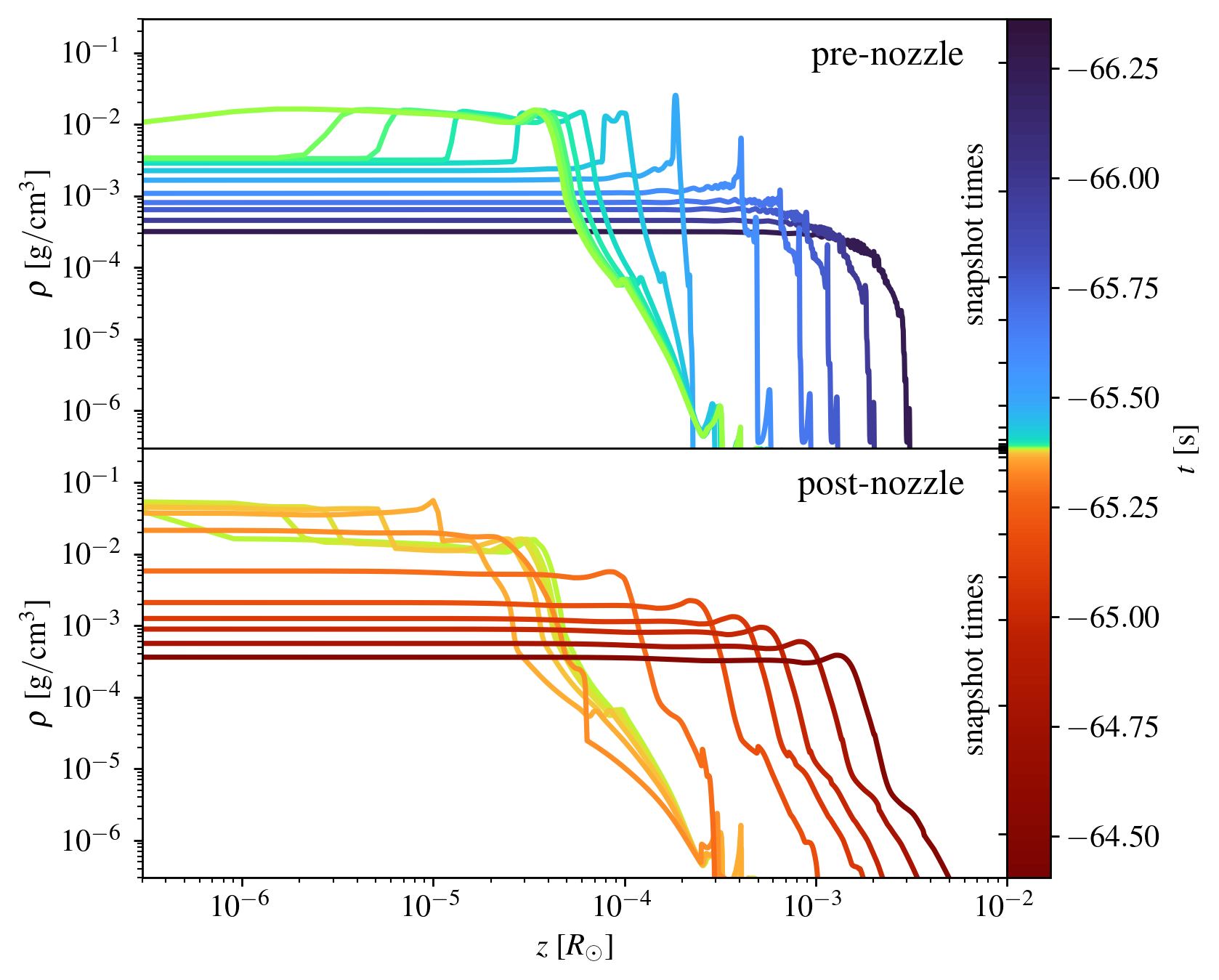}
    \caption{The density as a function of vertical height at 24 snapshots in the \textsc{Athena++} simulations for the fiducial stream slice with a \citetalias{Tomida+2013} EOS. We show 12 profiles before (top panel) and after (bottom panel) the nozzle. We indicate the time of each profile by its color in the color bar. The snapshots and the color bar are concentrated near the nozzle, where the evolution is most rapid. Before the nozzle, a shock forms and propagates towards the center of the stream slice, where it collides with the shock from the other side of the $xy$-plane. The resulting dissipation launches outward-propagating shocks that rapidly reach the surface of the stream slice and break out. At the nozzle, the vertical height of the stream is $10^{-4}~R_\odot$. }
    \label{fig:profile}
\end{figure}

In Figure~\ref{fig:profile}, we show the density profiles in the \texttt{Athena++} simulations at 17 snapshots for the fiducial stream slice with a \citetalias{Tomida+2013} EOS. Before the nozzle, a shock forms and propagates towards the center of the stream slice, where it collides with the shock from the other side of the $xy$-plane. The resulting dissipation launches outward-propagating shocks that rapidly reach the surface of the stream slice and break out, analogous to shock breakout in the initial disruption for deep encounters \citep{Carter&Luminet1983, Guillochon+2009, Yalinewich+2019}. The shock breakout removes material from the surface of the stream slice, producing a shallower outer density profile compared to before the nozzle.

\begin{figure}
    \centering
    \includegraphics[width=0.9\linewidth]{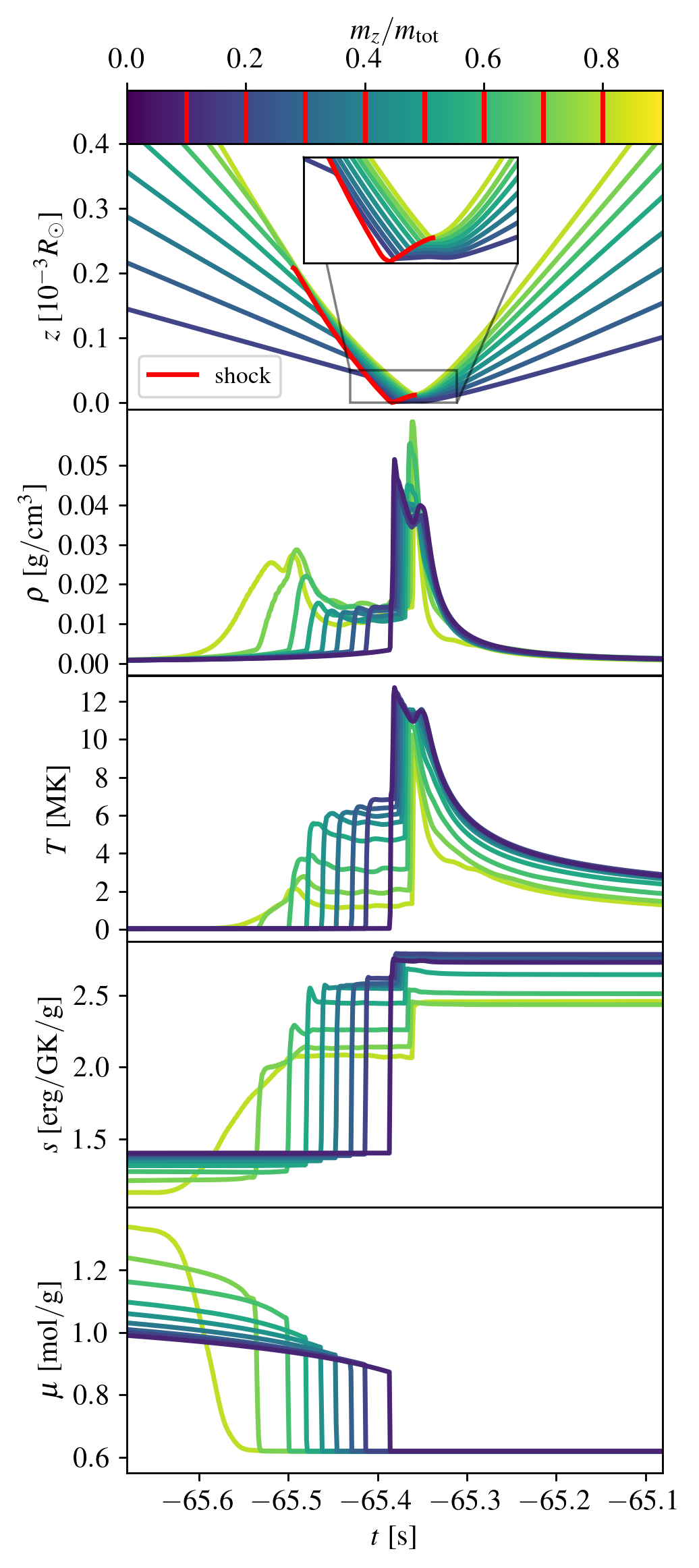}
    \caption{The vertical coordinate (first panel), density (second panel), temperature in megaKelvin (MK) (third panel), specific entropy in units erg/GK/g where GK is gigaKelvin (fourth panel), and mean molecular weight (fifth panel) at several Lagrangian mass coordinates as a function of time near the nozzle in the \textsc{Athena++} simulations for the fiducial stream slice with a \citetalias{Tomida+2013} EOS. In the first panel inset axes, we zoom in near the nozzle and indicate the shock front in red. We indicate the value of each mass coordinate by its color in the color bar above the first panel. The central density and temperature at maximum compression reach $0.05~{\rm g/cm^3}$ and $12~{\rm MK}$ respectively. Mass coordinates $> 0$ encounter the shock twice: first as it propagates inward before the nozzle and second as it propagates outward after the nozzle. Each shock encounter is associated with a nearly instantaneous increase in entropy. There is a similar net entropy increase at all mass coordinates due to opposite trends in shock frame velocity versus mass coordinate at each shock encounter. }
    \label{fig:lag}
\end{figure}

The nozzle shock can alternatively be analyzed from a Lagrangian perspective. In Figure~\ref{fig:lag}, we show the properties at several Lagrangian mass coordinates as a function of time near the nozzle in the \textsc{Athena++} simulations for the fiducial stream slice with a \citetalias{Tomida+2013} EOS. We interpolate the temperature, specific entropy, and mean molecular weight from the EOS table. The central density and temperature at maximum compression reach $0.05~{\rm g/cm^3}$ and $13~{\rm megaKelvin}$ (MK) respectively, compared to $3.4\times 10^{-3}~{\rm g/cm^3}$ and $42000~{\rm K}$ just before maximum compression.

At the shock front, the locations of different mass coordinates converge due to the discontinuity. Mass coordinates $\ne 0$ encounter the shock twice: first as it propagates inward before the nozzle and second as it propagates outward after the nozzle. Each shock encounter is associated with a nearly instantaneous increase in entropy. The fluid elements at every mass coordinate are ionized by the first shock encounter. This alters the shock jump conditions, allowing the compression ratio to exceed the usual factor 4 for strong shocks with adiabatic index $\gamma = 5/3$.

Fluid elements at larger mass coordinate have larger negative vertical velocities, but all fluid elements have smaller negative vertical velocities than the inward-propagating shock. Therefore, in the shock frame, fluid elements at larger mass coordinate have smaller positive velocities and longer shock-crossing times. A long shock-crossing time explains the more gradual increase in entropy during the first shock encounter for the fluid element at the largest mass coordinate. 

As the shock rebounds from the center of the stream slice, the other fluid elements are still collapsing with negative vertical velocities. Therefore, in the shock frame, fluid elements at larger mass coordinate have larger negative velocities. 

The range of shock Mach numbers across fluid elements is primarily due to the differences in shock frame velocities, since the differences in upstream sound speeds are relatively smaller. Therefore, at larger mass coordinates, there is less dissipation at the first shock encounter and more dissipation at the second shock encounter. These effects balance to give a similar net entropy increase at all mass coordinates. 

When the outward-propagating shock breaks out, it launches a pressure wave back towards the center of the stream slice, where it collides with the wave from the other side of the $xy$-plane, producing a second peak in the central density and temperature. However, these pressure waves never steepen into shocks, and thus do not contribute to the energy dissipation. 

\subsection{Energy dissipation at the nozzle}
\label{sec:ener_diss}

\begin{figure}
    \centering
    \includegraphics[width=\linewidth]{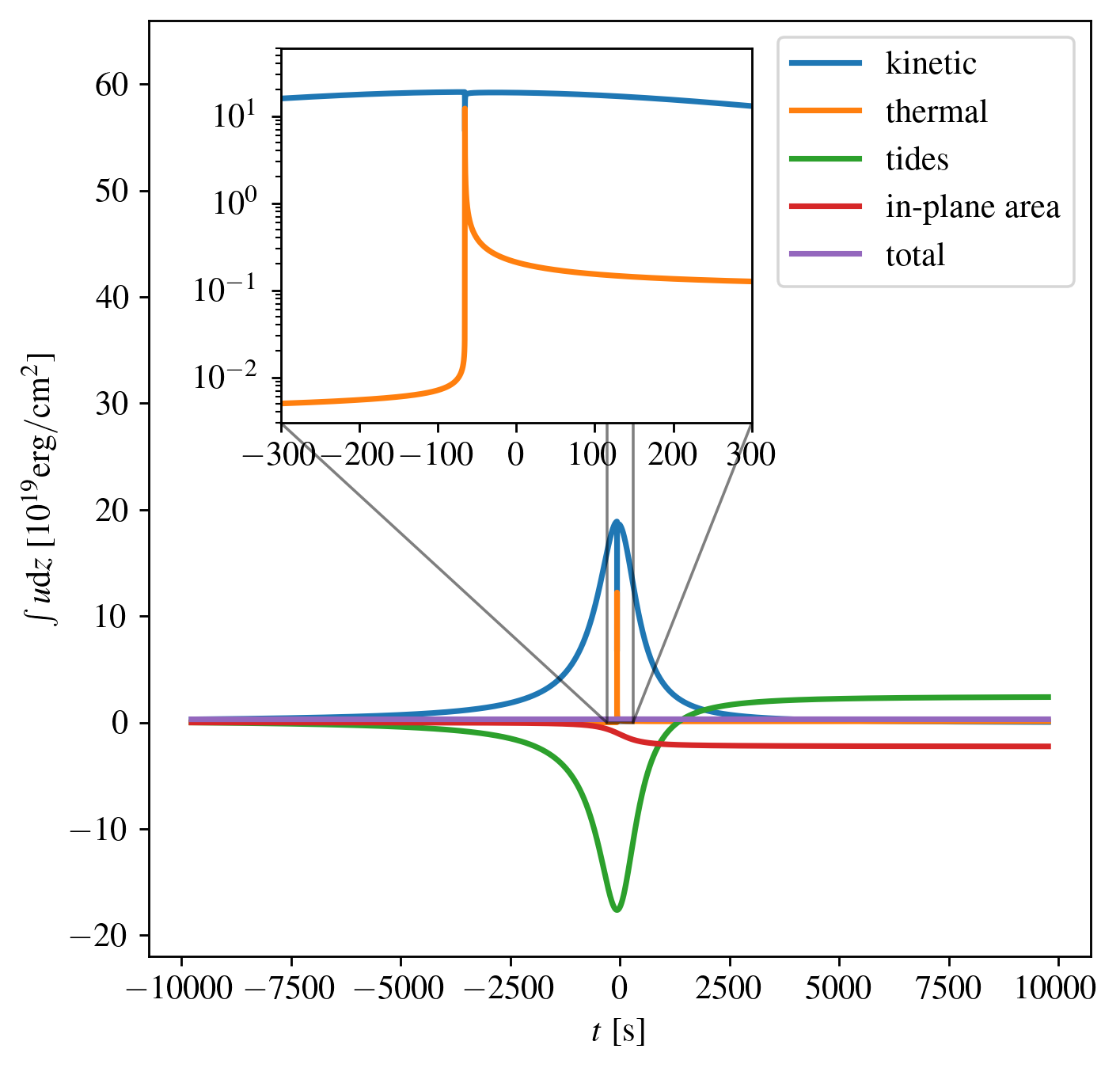}
    \caption{The contributions to the 1D integrated energy density as a function of time in the \textsc{Athena++} simulations for fiducial stream slice with a \citetalias{Tomida+2013} EOS. The contributions include kinetic energy (blue line), thermal energy (orange line), and source terms from tides (green line) and in-plane area evolution (red line). We also show their sum, the total energy (purple line). In the inset panel, we show a zoom-in of the evolution near the pericenter passage, with the $y$-axis on a log scale. At the nozzle, kinetic and thermal energy are exchanged by the shock, ultimately increasing the thermal energy by a factor of about $20$. }
    \label{fig:energy}
\end{figure}

The shock dissipates kinetic energy associated with the vertical compression in favor of thermal energy. In Figure~\ref{fig:energy}, we show contributions to the 1D integrated energy density as a function of time in the \textsc{Athena++} simulations for fiducial stream slice with a \citetalias{Tomida+2013} EOS. The contributions include kinetic energy, thermal energy, and source terms from tides and in-plane evolution. The kinetic and thermal energy contributions are $\int (1/2) \dd z \rho v_z^2$ and $\int \dd z e$ respectively. For each source term $\mathcal{S}_i$, the contribution is $\int \dd t \int \dd z \mathcal{S}_i$. Their sum, the total energy, is conserved to within one part in $10^6$, except at the shock where energy conservation is violated by a few percent due to operator splitting between the advection and source terms.

As the stream slice returns to pericenter, the tides do work to increase the kinetic energy. At the nozzle, kinetic and thermal energy are exchanged by the shock, ultimately increasing the thermal energy by a factor of about $20$. As the stream slice leaves pericenter, the tides do work to decrease the kinetic energy.

\begin{figure}
    \centering
    \includegraphics[width=\linewidth]{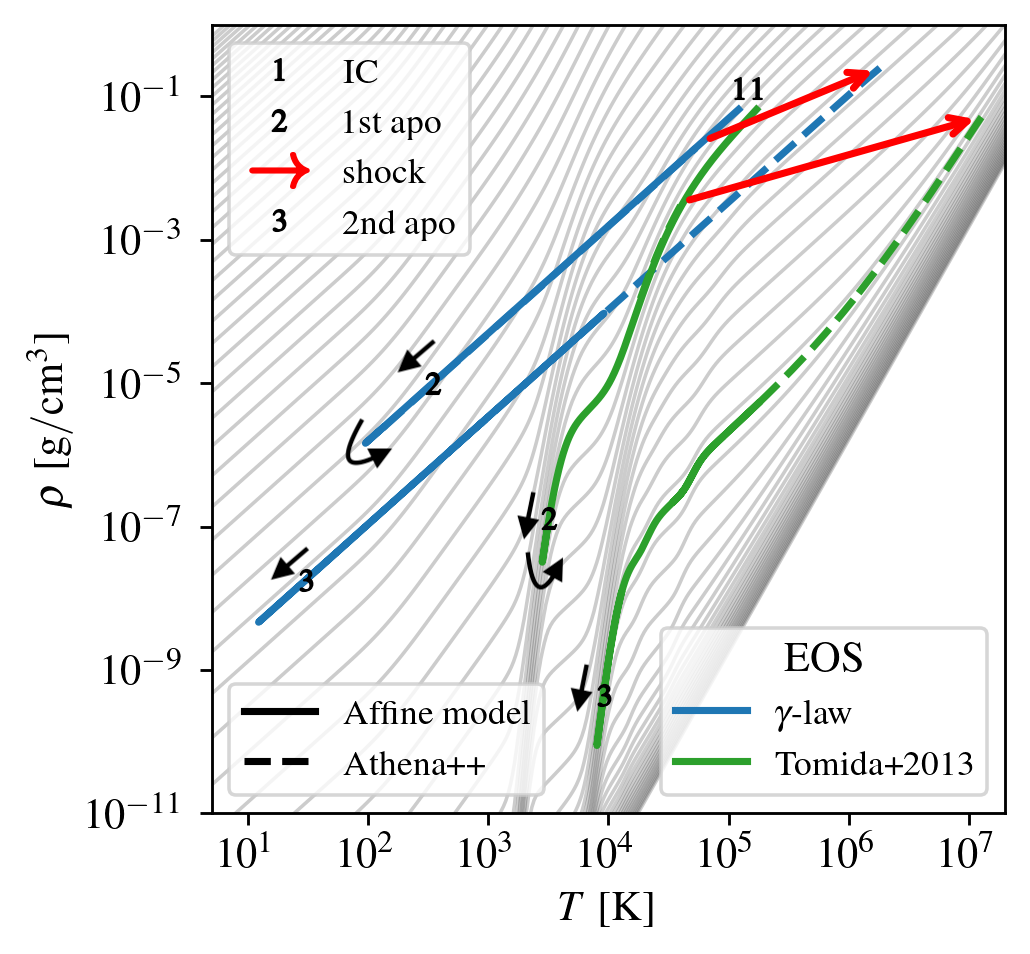}
    \caption{The evolution of the fluid element at the center of the fiducial stream slice in temperature-density phase space in the affine model (solid lines) and \textsc{Athena++} simulations (dashed lines). We show results for $\gamma$-law (blue line) and \citetalias{Tomida+2013} (green line) EOSs. For the \citetalias{Tomida+2013} EOS, we also show adiabats (grey lines). For the $\gamma$-law EOS, adiabats follow $\rho \propto T^{2/3}$. Starting from the initial conditions (IC; label 1), the stream slice evolves along an adiabat during the first orbit, expanding as it passes through apocenter (2) and compressing as it returns to the BH. At the nozzle, shock dissipation moves the debris stream to a higher-entropy adiabat (red arrows), on which it evolves during the second orbit, expanding as it passes through apocenter (3).}
    \label{fig:phasespace}
\end{figure}

In Figure~\ref{fig:phasespace}, we show the evolution of the fluid element at the center of the fiducial stream slice in temperature-density phase space in the affine model and \textsc{Athena++} simulations. In this phase space, the logarithmic slope is equal to the effective adiabatic index $\Gamma_3 - 1$. For the $\gamma$-law EOS, adiabats follow $\rho \propto T^{2/3}$. The initial temperatures differ for the different EOSs because the initial density and pressure are fixed. 

The stream slice evolves along an adiabat during the first orbit, expanding as it passes through apocenter and compressing as it returns to the BH. At the nozzle, shock dissipation moves the debris stream to a higher-entropy adiabat, on which it evolves during the second orbit. The nozzle shock is stronger for the \citetalias{Tomida+2013} EOS, so the entropy increases by a larger amount compared to the $\gamma$-law EOS. 

During chemical processes, the adiabats are steeper, so the temperature stays nearly constant over a large range in density. For the \citetalias{Tomida+2013} EOS, ${\rm H_2}$ formation maintains a gas temperature $\gtrsim 3000~{\rm K}$ on the first orbit and H recombination maintains a gas temperature $\gtrsim 6000~{\rm K}$ on the second orbit. For the \citetalias{Tomida+2013} EOS, the pressure at the nozzle has a significant contribution from radiation pressure, with $\beta_{\rm rad} \simeq 0.44$.

\begin{figure}
    \centering
    \includegraphics[width=\linewidth]{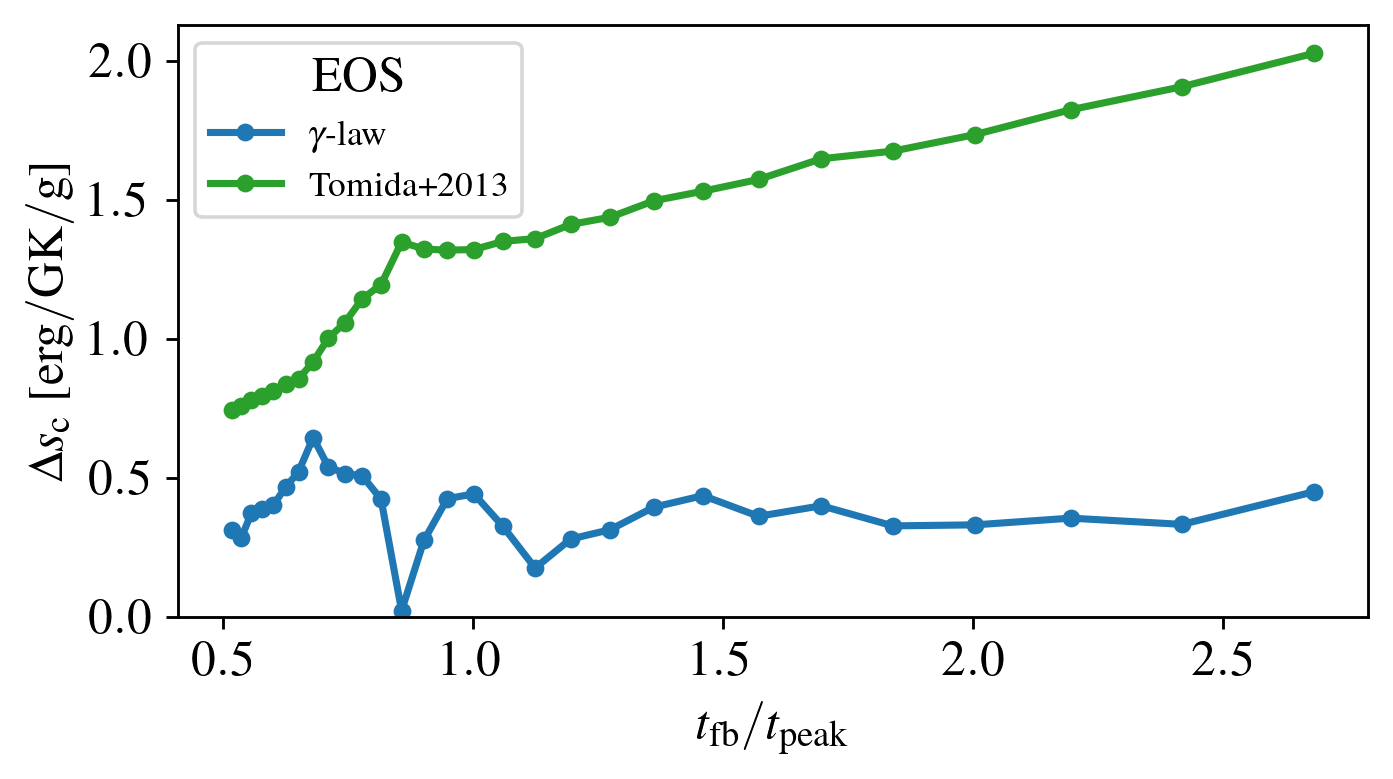}
    \caption{The change in specific entropy for the fluid element at the center of the stream slice in the \textsc{Athena++} simulations as a function of stream slice fallback time relative to peak. We show results for $\gamma$-law (blue lines) and \citetalias{Tomida+2013} (green lines) EOSs. There is significantly more dissipation for the latter EOS due to pre-nozzle stream expansion driven by H recombination and ${\rm H_2}$ formation. }
    \label{fig:max}
\end{figure}

We quantify the dissipation using the change in specific entropy at the center of the stream slice $\Delta s_{\rm c}$. In Figure~\ref{fig:max}, we show $\Delta s_{\rm c}$ in the \textsc{Athena++} simulations for all stream slices. For the \citetalias{Tomida+2013} EOS, we interpolate the specific entropy from the EOS table. For the $\gamma$-law EOS, the specific entropy is $s = c_V \ln p / \rho^\gamma$, where $c_V = (\gamma - 1)^{-1} k_{\rm B} / \mu m_{\rm p}$ is the specific heat at constant volume. 

The entropy increase for the \citetalias{Tomida+2013} EOS ranges from $0.7$ to $2~{\rm erg/GK/g}$ where GK is gigaKelvin, while for the $\gamma$-law EOS it is $\lesssim 0.5~{\rm erg/GK/g}$. With the \citetalias{Tomida+2013} EOS, dissipation increases quickly at fallback times $\lesssim 0.85 t_{\rm peak}$ because those stream slices enter the ballistic regime while H recombination is ongoing and would otherwise continue to drive stream expansion. At later fallback times, the trend continues with a shallower slope due to ongoing ${\rm H_2}$ formation. 

With the $\gamma$-law EOS, the amount of dissipation is sensitive to the phase of the quadrupolar oscillations at the transition to the ballistic regime. For the stream slice at $t_{\rm fb}/t_{\rm peak} \simeq 0.85$, the vertical width happens to be at a trough in the oscillation at the transition, reducing the energy available to dissipate at the nozzle. As a result, the change in specific entropy is an order-of-magnitude smaller than the other stream slices. 

The strength of the shock can alternatively be described by the Mach number. Using the shock jump conditions with a $\gamma=5/3$ EOS, the change in specific entropy is related to the Mach number in the strong shock limit by,
\begin{equation}\begin{split}
	\Delta s \approx\ & c_V \ln \left( \frac{2 \gamma (\gamma - 1)^{\gamma} \mathcal{M}_{\rm sh}^2}{(\gamma + 1)^{\gamma + 1}} \right)\\
    \simeq\ & (0.50 \ln \mathcal{M}_{\rm sh} - 0.52)~{\rm erg/GK/g}
\end{split}\end{equation}
With a $\gamma$-law EOS, the typical Mach number of the nozzle shock ranges from $5$ to $10$. 

With a \citetalias{Tomida+2013} EOS, this calculation no longer holds because the gas is ionized at the shock. The relationship between $\Delta s_{\rm c}$ and $\mathcal{M}_{\rm sh}$ depends on the fraction of energy that goes into ionization, and thus on the pre-shock density. However, we can estimate the Mach numbers for the fiducial stream slice using the data from Figure~\ref{fig:lag}, which gives values around $60$.

\subsection{Affine model evolution on the second orbit}
\label{sec:second_orb}

The orbital planes of all fluid elements intersect along a line. When the intersection line crosses the orbit, the fluid elements are ballistically focused towards the $xy$-plane until enough pressure builds up to reverse the collapse \citep{Carter&Luminet1983, Stone+2013}. The intersection line passes through the BH and lies parallel to the orbital velocity at the point in the first orbit where $\partial_t H$ changes sign \citep{Luminet&Marck1985}. 

\begin{figure*}
    \centering
    \includegraphics[width=0.67\linewidth]{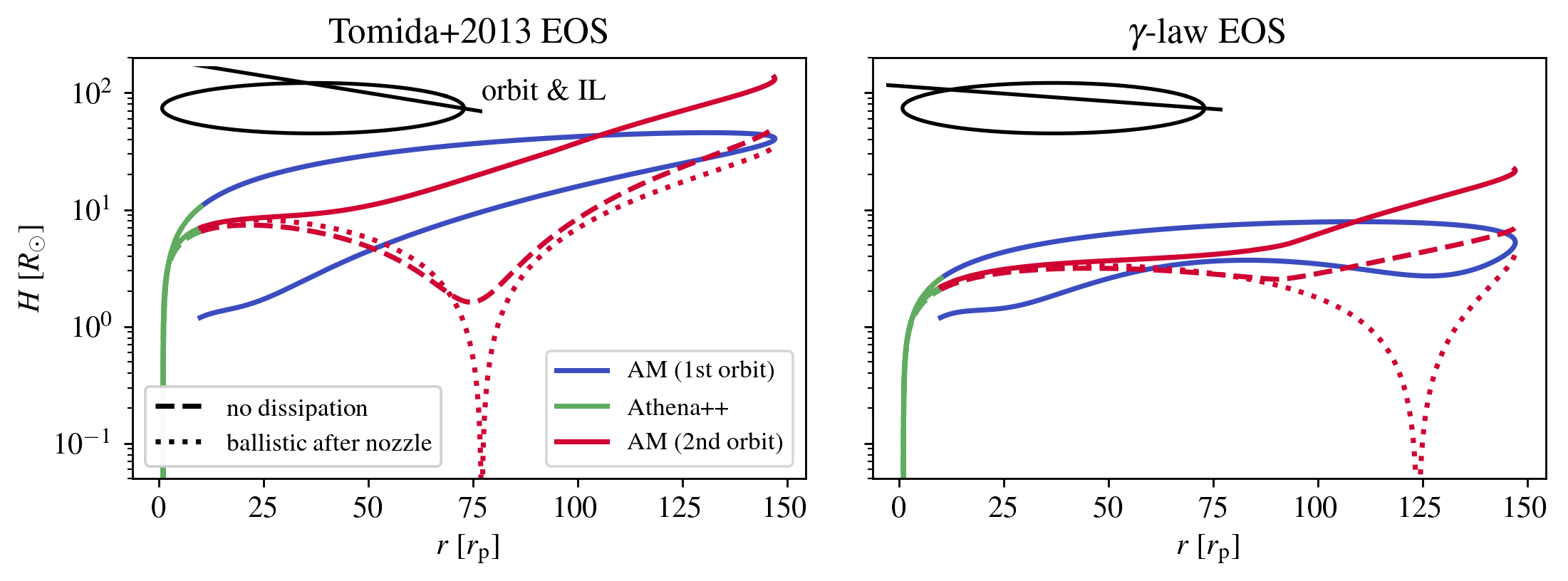}
    \caption{The vertical stream width as a function of radius for the fiducial stream slice with a \citetalias{Tomida+2013} (left panel) and $\gamma$-law (right panel) EOS. We show the affine model evolution on the first orbit (blue lines), the \textsc{Athena++} evolution during pericenter passage (green lines), and the affine model evolution on the second orbit (red lines). For the latter, we show 3 versions: default (solid lines), calculated with the second affine model; no dissipation (dashed lines), calculated using the first affine model; and purely ballistic (dotted lines), calculated by removing pressure terms from the second affine model. Where the intersection line (IL) crosses the orbit (top left corner of each panel), there are focal points: one near pericenter and one at $77~r_{\rm p}$ (resp. $124~r_{\rm p}$) for the \citetalias{Tomida+2013} (resp. $\gamma$-law) EOS. Pressure gradients, enhanced by dissipation at the nozzle, prevent collapse at the second focal point that would otherwise occur. Because pressure gradients dominate over tides for portions of the second orbit, the stream widths depend on the amount of dissipation at the nozzle.  }
    \label{fig:orbits}
\end{figure*}

This construction always produces two focal points: one close to pericenter $\le 2 r_{\rm p}$ responsible for the extreme vertical compression at the nozzle and one at larger radii \citep{Bonnerot&Lu2022}. At the nozzle, the intersection line precesses in the direction of motion because the vertical pressure gradient torques the inclined orbits of the fluid elements \citep{Bonnerot&Lu2022}. In Figure~\ref{fig:orbits}, we show the post-nozzle orientation of the intersection line for the fiducial stream slice. The intersection line crosses the orbit at $77~r_{\rm p}$ (resp. $124~r_{\rm p}$) with a \citetalias{Tomida+2013} (resp. $\gamma$-law) EOS. 

We also show the vertical stream width as a function of radius, including the affine model evolution on the first orbit, the \textsc{Athena++} evolution during the pericenter passage, and the affine model evolution on the second orbit. For the latter, we show 3 versions of the evolution: with dissipation at the nozzle, calculated with the second affine model; without dissipation, calculated with the first affine model; and purely ballistic, calculated by removing pressure terms from the second affine model. These are effectively a series of decreasing pressure in the post-nozzle stream.

In the purely ballistic evolution, the vertical stream height goes to zero at the second focal point, since there are no pressure gradients to resist the collapse. When pressure is included, but no dissipation, there is modest vertical compression to about $2~R_\odot$ with a \citetalias{Tomida+2013} EOS. The compression is less extreme than at the nozzle because tides are weaker at larger radii. With a $\gamma$-law EOS, the second focal point occurs even further out, so pressure gradients almost entirely prevent the collapse. 

When dissipation is included, the enhanced pressure gradients completely resist the collapse for both EOSs and drive significant stream expansion. With a \citetalias{Tomida+2013} EOS, stream expansion on the second orbit is additionally augmented by H recombination. Because pressure gradients dominate over tides for portions of the second orbit, the stream widths depend on the amount of dissipation at the nozzle.

Other stream slices have similar post-nozzle evolution, except for the most bound stream slices with a $\gamma$-law EOS. These have equilibrium points before apocenter, setting up a second focal point before the nozzle and producing extra density peaks in Figure~\ref{fig:rho}. With a \citetalias{Tomida+2013} EOS, the stronger pressure gradients move the equilibrium point after apocenter. This occurs even when the stream slice is initially ballistic, because the ratio of pressure gradient to tidal forces tends to increase during H recombination (App.~\ref{sec:toy_model}).

During ballistic evolution, there are also in-plane focal points where the debris stream collapses towards its centerline \citep{Kochanek1994}. On the second orbit, there is an in-plane focal point at $94~r_{\rm p}$ (resp. $93~r_{\rm p}$) with a \citetalias{Tomida+2013} (resp. $\gamma$-law) EOS where the stream collapses to an in-plane width $0.6~R_\odot$ ($0.2~R_\odot$). The stream collapses in the plane despite the pressure gradient force because the effective in-plane tidal force on the second orbit is enhanced by strong in-plane shear (Eq.~\ref{eq:yp_eom}).

\subsection{Conditions at the self-intersection}
\label{sec:si_cond}

For a non-spinning BH, the argument of periapse $\omega$ advances per orbit by approximately $\delta \omega = 6\pi (1 + e)^{-1} (r_{\rm g} / r_{\rm p})$, where $r_{\rm g} \equiv G M_\bullet / c^2$ is the BH gravitational radius. Treating the orbit as an ellipse with instantaneous precession at pericenter \citep{Dai+2015}, the orbit self-intersects at radius and angle
\begin{align}
    r_{\rm si} =\ & \frac{(1 + e) r_{\rm p}}{1 - e \cos (\delta \omega / 2)} \label{eq:rsi} \\
    \cos \Theta_{\rm si} =\ & \frac{1 - 2 \cos (\delta \omega / 2) e + \cos (\delta \omega) e^2}{1 - 2 \cos (\delta \omega / 2) e + e^2} \label{eq:thsi}
\end{align}
where $\Theta_{\rm si}$ is the angle between the incoming and outgoing stream velocities (Fig.~\ref{fig:diagram}). These expressions match the exact solutions given by integrating the geodesic equation in the Schwarzschild spacetime for TDEs with $r_{\rm g} / r_{\rm p} \lesssim 0.1$ \citep{Lu&Bonnerot2020}.

The tangential and radial components of the two colliding stream velocities are parallel and antiparallel respectively, so only the kinetic energy associated with the radial motion is dissipated in the self-intersection. To lowest order in $r_{\rm g}/r_{\rm p}$ and $1-e$, the radial velocity at the self-intersection point is
\begin{equation}\begin{split}
    \frac{v_{r, {\rm si}}}{c} =\ & \left( \frac{2}{r_{\rm si}/r_{\rm g}} - \frac{1 - e}{r_{\rm p}/r_{\rm g}} \right)^{1/2} \sin ( \Theta_{\rm si} / 2 )\\
    \approx\ & \frac{3\pi}{\sqrt{8}} \left( \frac{r_{\rm g}}{r_{\rm p}} \right)^{3/2} \simeq 0.01 \beta^{3/2} Q_6
    \label{eq:vrsi}
\end{split}\end{equation}
where $Q_6 \equiv Q / 10^6$.

Using Kepler's 3rd law $\varepsilon = -(1/2)(2\pi GM_\bullet / t_{\rm fb}^2)^{2/3}$, the specific orbital energy changes on a timescale $\varepsilon / |\dot{\varepsilon}| = (3/2)t_{\rm fb}$. Approximating the orbits as radial and parabolic, the difference in fallback time between the incoming and outgoing streams at self-intersection is $\Delta t_{\rm fb} \approx (2/3)(2 r_{\rm si}^3/GM_\bullet)^{1/2}$. For our TDE parameters, we have $\varepsilon / |\dot{\varepsilon}| \gg \Delta t_{\rm fb}$ for $t_{\rm fb} \gg (4/9) (2 r_{\rm si}^3 / GM_\bullet)^{1/2} \simeq 2~{\rm day}$. This is before even the most bound material returns to the BH (Fig.~\ref{fig:mdot}), so we ignore the difference in orbital properties between the incoming and outside stream. 

\begin{figure}
    \centering
    \includegraphics[width=\linewidth]{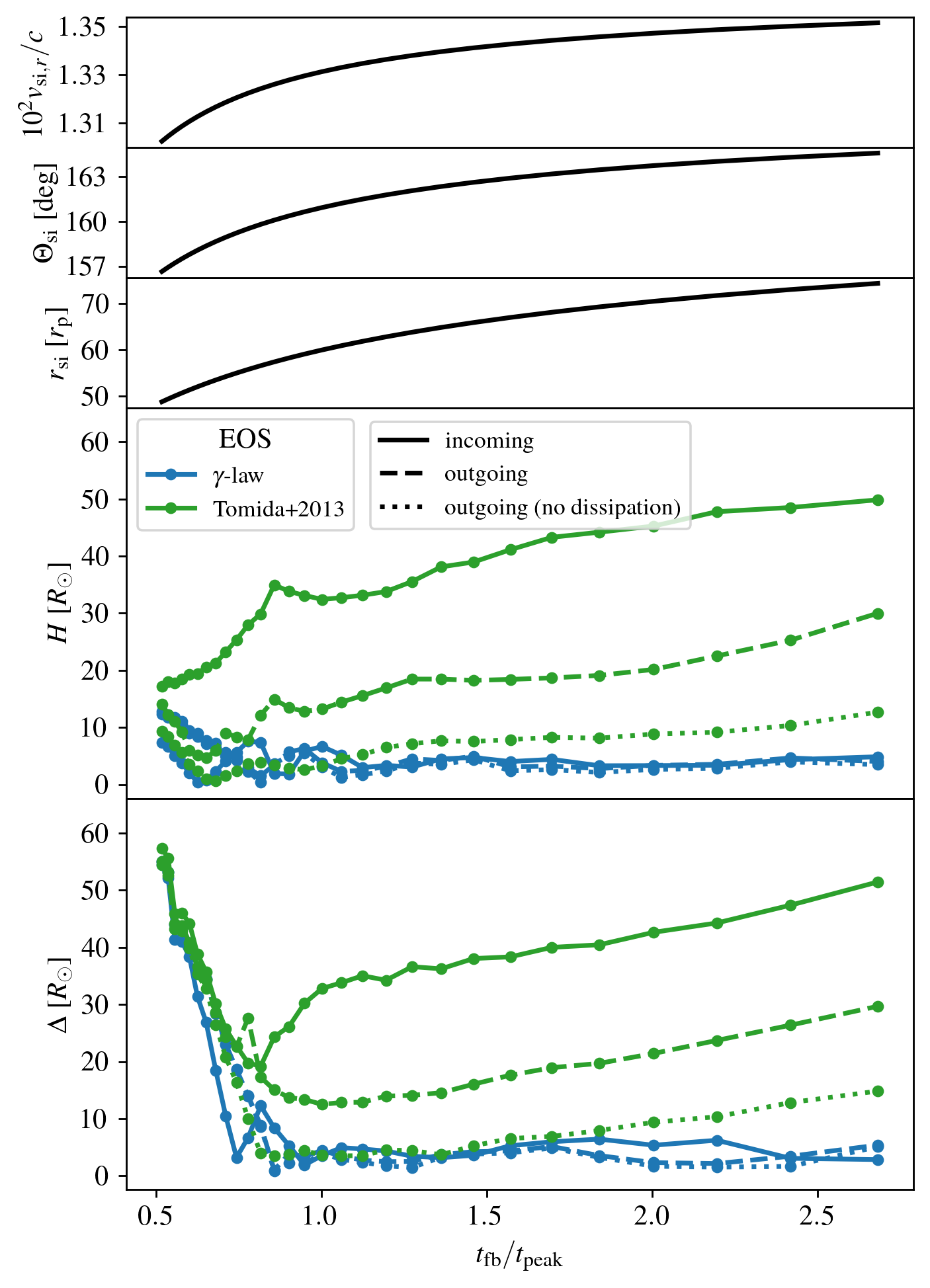}
    \caption{The radial velocity (first panel, Eq.~\ref{eq:vrsi}), intersection angle (second panel, Eq.~\ref{eq:thsi}), radial coordinate (third panel, Eq.~\ref{eq:rsi}), stream vertical widths (fourth panel), and stream transverse widths (fifth panel) at the self-intersection for all stream slices as a function of by their fallback time relative to peak. The incoming (solid lines) and outgoing (dashed lines) stream widths are calculated using the first and second affine models respectively. We also show the outgoing stream widths in the case of zero dissipation at the nozzle (dotted lines), calculated using the first affine model. We show results with $\gamma$-law (blue line) and \citetalias{Tomida+2013} (green line) EOSs. With a $\gamma$-law EOS, the incoming and outgoing stream widths are similar, regardless of dissipation at the nozzle. With a \citetalias{Tomida+2013} EOS, the incoming stream width is larger than the outgoing stream width by a factor of a few, and the outgoing stream width is sensitive to dissipation at the nozzle. We may under-estimate the in-plane widths of the outgoing stream, since we do not include differential apsidal precession (Sec.~\ref{sec:GR}). }
    \label{fig:si}
\end{figure}

In Figure~\ref{fig:si}, we show the radial velocity, intersection angle, radial coordinate, and incoming and outgoing stream widths at the self-intersection. The incoming and outgoing stream widths are calculated using the first and second affine models respectively. We also show the outgoing stream widths in the case of zero dissipation at the nozzle, calculated using the first affine model.

The typical self-intersection properties are $v_{{\rm si}, r} \simeq 0.01c$, $\Theta_{\rm si} \simeq 160^\circ$, and $r_{\rm si} \simeq 50-70~r_{\rm p}$. The typical stream widths at self-intersection are tens of (resp. a few) solar radii with a \citetalias{Tomida+2013} (resp. $\gamma$-law) EOS. However, we may underestimate the in-plane stream width because we do not include the effect of differential apsidal precession (Sec.~\ref{sec:GR}). 

The most bound stream slices have larger in-plane stream widths due to their larger in-plane widths in the affine model initial conditions (Sec.~\ref{sec:init_dis_res}). With a $\gamma$-law EOS, the incoming and outgoing stream widths are similar, regardless of dissipation at the nozzle, similar to \citet{Bonnerot&Lu2022}. This is possible, despite our results in Section~\ref{sec:second_orb}, because deviations from ballistic evolution only appear near the second focal point, which lies far outside self-intersection radius for the $\gamma$-law EOS (Fig.~\ref{fig:orbits}).

With a \citetalias{Tomida+2013} EOS, the second focal point occurs close to the self-intersection, slowing the expansion of the outgoing stream as it approaches the self-intersection. As a result, the incoming stream width is larger than the outgoing stream width by a factor of a few, similar to \citet{Kochanek1994}, who also include H recombination. This also makes the outgoing stream height sensitive to dissipation at the nozzle. 

The discrepancy between incoming and outgoing stream heights suggests that for a realistic EOS, only a fraction of the incoming debris stream participates in the self-intersection, substantially reducing the energy available to dissipate at the self-intersection shock.

\subsection{Convergence test}
\label{sec:conv_test}

\begin{figure}
    \centering
    \includegraphics[width=\linewidth]{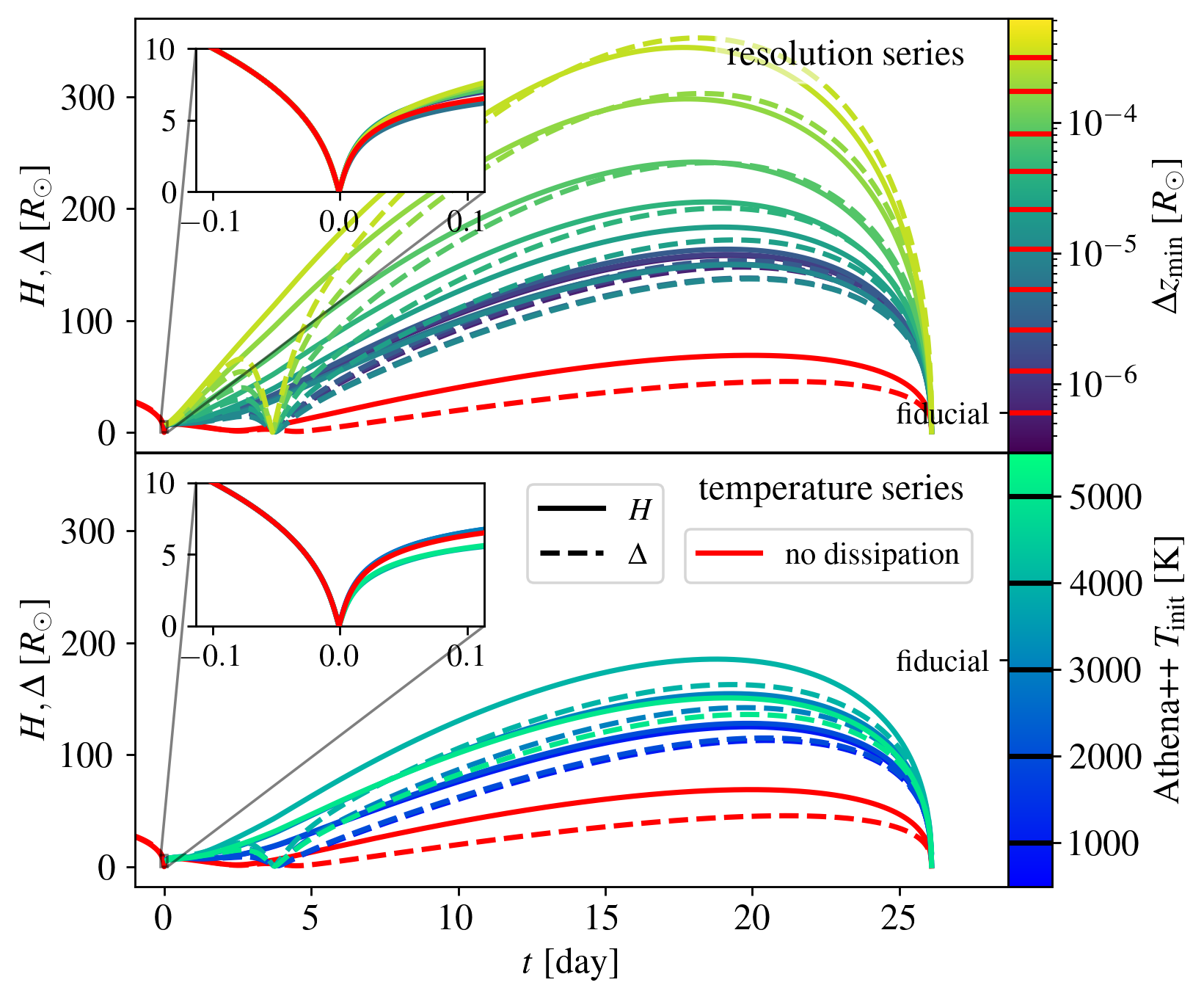}
    \caption{The vertical (solid lines) and in-plane (dashed lines) stream widths as a function of time for the fiducial stream slice with the \citetalias{Tomida+2013} EOS, varying the resolution at the center of the stream slice (top panel) and the initial temperature (bottom panel) in the \textsc{Athena++} simulations. We show the evolution at the nozzle (inset panels) and on second orbit (main panels) using data from the \textsc{Athena++} simulations and the second affine model respectively. We also show the evolution with no dissipation using data from the first affine model (red lines). We indicate each resolution and initial temperature by its color in the color bar of the corresponding panel, labeling the fiducial value with an annotation. The stream widths on the second orbit are sensitive to the degree of dissipation at the nozzle, with more dissipation producing larger stream widths. As the resolution increases, there is less numerical dissipation and the stream widths decrease asymptotically towards their physical values, converging for $\Delta z_{\rm min} \lesssim 5\times 10^{-6}~R_\odot$. As the initial temperature increases, the Mach number of the nozzle shock decreases, reducing dissipation and increasing stream widths. These results demonstrate that in order to accurately predict the outgoing stream properties at self-intersection, one must resolve the nozzle shock and accurately model the thermodynamics of the stream. }
    \label{fig:series}
\end{figure}

In Figure~\ref{fig:series}, we show the stream widths as a function of time for the fiducial stream slice with a \citetalias{Tomida+2013} EOS over a series of \textsc{Athena++} simulations with different resolution and initial temperature. We show the evolution at the nozzle and on second orbit using data from the \textsc{Athena++} simulations and the second affine model respectively. We also show the evolution with no dissipation using data from the first affine model, using the fiducial initial temperature. 

We decrease the central resolution compared to fiducial value $7.0\times 10^{-7}~R_\odot$ by changing the size ratio between neighboring cells on our logarithmic grid, but we leave the number of cells fixed at $2^{14} = 16384$. The lowest central resolution we consider is $3.1 \times 10^{-4}~R_\odot$, which is larger than our fiducial resolution by a factor $520$ and certainly does not resolve the nozzle shock. 

The stream widths on the second orbit are sensitive to the amount of dissipation at the nozzle, with more dissipation producing larger stream widths. As the resolution increases, there is less numerical dissipation and the stream widths decrease asymptotically towards their physical values, converging for $\Delta z_{\rm min} \lesssim 5\times 10^{-6}~R_\odot$. Near the nozzle, the differences in stream widths are less apparent because the evolution there is dominated by tides.

We also vary the initial temperature of the stream slice at the start of the \textsc{Athena++} simulation. We use values 1000, 2000, 3000, 4000, and $5000~{\rm K}$, which cover above and below the value of $3139~{\rm K}$ in our fiducial simulation. As the initial temperature increases, the Mach number of the nozzle shock decreases, reducing dissipation and increasing stream widths. These results demonstrate that in order to accurately predict the outgoing stream properties at self-intersection, one must resolve the nozzle shock and accurately model the thermodynamics of the stream.

%% file: sections/04_discussion.tex
\subsection{Comparison to previous work}
\label{sec:comp_prev}

Only a few previous works on the debris stream evolution include chemical processes \citep{Kochanek1994, Kasen&RamirezRuiz2010, Coughlin2023, Steinberg&Stone2024}, and even then they typically only include H recombination. Our results demonstrate that H recombination and ${\rm H_2}$ formation have a significant impact on the debris stream, the nozzle shock, and stream self-intersection. They are essential to include in future work on debris stream dynamics and early circularization.

In contrast to previous work, we find that dissipation at the nozzle produces neither rapid stream expansion \citep{Lee&Kim1996, Ayal+2000, Guillochon+2014, Shiokawa+2015, Ryu+2023, Steinberg&Stone2024} nor minimal stream expansion \citep{Bonnerot&Lu2022, FitzHu+2025, Kubli+2025}. Instead, dissipation at the nozzle broadens the post-nozzle stream by a factor of a few at the self-intersection point, and a factor of several at apocenter.

Previous work shows that the post-nozzle stream width is not converged at the resolution of most 3D simulations \citep{Price+2024, Huang+2024}, so 3D simulations that find rapid stream expansion may be dominated by numerical dissipation at the nozzle. In the BH frame, the kinetic energy at the nozzle is dominated by the orbital motion, so even a little numerical dissipation can dramatically overestimate the physical value \citep{Bonnerot&Lu2022}.

On the other hand, simulations that find minimal stream expansion do not include chemical processes. In our simulations, chemical processes modify the timing of the second focal point such that pressure gradient forces can drive stream expansion before self-intersection (Fig.~\ref{fig:orbits}). The importance of pressure in the post-nozzle stream was acknowledged by \citet{Bonnerot&Lu2022} but not analyzed in detail.

This result depends sensitively on the dynamics during the first orbit, which sets the initial location of the second focal point, and on the compression and dissipation at the nozzle, which determines the amount by which the intersection line precesses. Therefore, the post-nozzle stream evolution may need to be analyzed for each new TDE setup on a case-by-case basis. In \citet{Bonnerot&Lu2022}, the second focal point is delayed until the end of the second orbit as the stream slice returns to pericenter. The difference in intersection line orientations may result from our higher penetration factor. 

\subsection{Implications for 3D simulations}
\label{sec:3d_imp}

At pericenter, the orbital velocity is $v_{\rm orb} / c = (1+e)^{1/2} (r_{\rm g}/r_{\rm p})^{1/2} \simeq 0.3 \beta^{1/2} Q_6^{1/3}$. The nozzle shock takes place over $\Delta t_{\rm noz} \simeq 1~{\rm s}$ (Fig.~\ref{fig:evol}), or equivalently, over a distance $\Delta x'_{\rm noz} = v_{\rm orb} \Delta t_{\rm noz} \simeq 0.1~R_\odot$. In the radial and vertical directions, the debris stream gets compressed to $\Delta y'_{\rm noz} \simeq 3~R_\odot$ and $\Delta z_{\rm noz} \simeq 10^{-4}~R_\odot$ respectively. Because $\Delta z_{\rm noz} \ll \Delta y'_{\rm noz}, \Delta x'_{\rm noz}$, the nozzle shock occurs in a 2D pancake-shaped region.

Several Eulerian finite-volume codes have been used in 3D global TDE simulations, including \textsc{FLASH} \citep{Fryxell+2000} e.g. \citet{Guillochon&RamirezRuiz2013}; \textsc{KORAL} \citep{Sadowski+2013} e.g. \citet{Sadowski+2016}; \textsc{Athena++} e.g. \citet{Huang+2024}; and \textsc{H-AMR} \citep{Liska+2022} e.g. \citet{Andalman+2022}. To date, no such simulation has fully resolved the vertical compression at the nozzle, even with static/adaptive mesh refinement (SMR/AMR).

Eulerian codes also struggle with high Mach flows because the internal energy is a small fraction of the total energy, so small truncation errors in the total energy can introduce large errors in the internal energy \citep{Ryu+1993, Bryan+1995, Feng+2004}. In the debris stream, the gas can reach Mach numbers $\gtrsim 10^4$ in the BH frame, even with ${\rm H_2}$ formation which keeps the gas temperature $\gtrsim 3000~{\rm K}$. Our convergence tests (Sec.~\ref{sec:conv_test}) indicate that one cannot get around this issue by increasing the gas temperature by-hand without affecting the accuracy of the stream widths on the second orbit.

Another approach is to use a finite-mass SPH code such as {\sc phantom} e.g. \citet{Bonnerot+2016} and \textsc{SPH-EXA} \citep{Cabezon+2025} e.g. \citet{Kubli+2025}. For a mass fallback rate $\dot{M}_{\rm fb} \sim M_\odot/{\rm yr}$, the mass in the nozzle region is $\dot{M}_{\rm fb} \Delta t_{\rm noz} \simeq 3 \times 10^{-8}~M_\odot$. To resolve the nozzle region with $10$ SPH particles across each dimension, one requires $\gtrsim 3\times 10^{10}$ SPH particles, which is consistent with the SPH particle numbers required for the post-nozzle stream width to converge \citep{Kubli+2025}.

There are also hybrid techniques that combine the advantages of explicit advection and Riemann-solver-based fluxes. These include moving mesh codes like \textsc{AREPO} \citep{Springel2010} e.g. \citet{Goicovic+2019} and \textsc{RICH} \citep{Yalinewich+2015} e.g. \citet{Steinberg&Stone2024} and the meshless finite-mass and finite-volume code \textsc{GIZMO} \citep{Hopkins2015} e.g. \citet{Pacuraru+2025}. These methods are promising for TDE simulations but are relatively new, and more work is needed to characterize their behavior in extreme shear and compression flows like at the nozzle.

\subsection{In-plane gradients at the nozzle}
\label{sec:plane_grad}

Our 1D treatment of the nozzle shock cannot capture the effect of in-plane pressure gradients. However, we expect them to have a small effect on the dynamical evolution of the debris. Near pericenter, strong tidal torques align the stream slice normal with the orbital velocity, so $\vu{y}'$ and $\vu{x}'$ are oriented radially and azimuthally respectively.

Pressure gradients in the $\vu{y}'$-direction are radial, so they cannot change the angular momentum. However, they might induce a spread in orbital energies if they produce a significant change in velocity relative to the orbital velocity. Using values for the fiducial stream slice at the nozzle (Fig.~\ref{fig:lag}), the relative change in velocity is
\begin{equation}\begin{split}
	\frac{\Delta v_{y'}}{v_{\rm orb}} \sim\ & \frac{1}{\rho} \frac{p}{\Delta} \frac{\Delta t_{\rm noz}}{v_{\rm orb}}\\
    \sim\ & 10^{-6} \left(\frac{p}{10^{14}{\rm erg/cm^3}}\right) \left(\frac{\Delta t_{\rm noz}}{1{\rm s}}\right)\left(\frac{\rho}{0.05{\rm g/cm^3}}\right)^{-1}\\
    \ & \left( \frac{\Delta y'_{\rm noz}}{3R_\odot}\right)^{-1} \left( \frac{v_{\rm orb}}{0.3 c}\right)^{-1}
\end{split}\end{equation}
This is a small number, so pressure gradients in the $\vu{y}'$-direction are not important for the dynamical evolution of the debris.

It is most convenient to treat the $\vu{x}'$-direction by shifting from the orbital frame to the BH frame. Time derivatives in the orbital and BH frames are related by $\partial_t|_{\rm orb} = \partial_t|_{\rm BH} + v_{\rm orb} \partial_{x'}$. We expect time derivatives in the BH frame to be much smaller than those in the orbital frame because the timescale of seconds spent at the nozzle is much shorter than the timescale of days over which the properties of the fallback change. Therefore, we ignore the term $\partial_t |_{\rm BH}$.

We estimate the change in orbital velocity associated with the pressure gradient force as
\begin{equation}\begin{split}
	\frac{\Delta v_{x'}}{v_{\rm orb}} \sim\ & \frac{1}{\rho} \eval{\pdv{p}{x'}}_{\rm BH} \frac{\Delta t_{\rm noz}}{v_{\rm orb}} \sim \frac{1}{\rho} \eval{\pdv{p}{t}}_{\rm orb} \frac{\Delta t_{\rm noz}}{v_{\rm orb}^2} \sim \frac{p}{\rho v_{\rm orb}^2}\\
	\sim\ & 10^{-5} \left(\frac{p}{10^{14}{\rm erg/cm^3}}\right)\left(\frac{\rho}{0.05{\rm g/cm^3}} \right)^{-1}\left( \frac{v_{\rm orb}}{0.3c} \right)^{-2}
\end{split}\end{equation}
This is a small number, so pressure gradients in the $\vu{x}'$-direction are not important for the dynamical evolution of the debris. The net change in orbital velocity would be even smaller for a more sophisticated calculation, because force contributions before and after the nozzle would partially cancel out.

We can also discount the possibility that the nozzle shock launches pressure waves in the $\vu{x}'$-direction into the pre-shocked, because the sound speed is much smaller than the orbital velocity at pericenter \citep{Bonnerot&Lu2022}. We conclude that the in-plane evolution at the nozzle is almost entirely ballistic, consistent with the assumption in our \textsc{Athena++} simulations. Fundamentally, this is because the energy budget at the nozzle is dominated by kinetic energy associated with the orbital motion. At most, the nozzle shock can dissipate all of the kinetic energy associated with the vertical motion, which is only a small fraction of the orbital kinetic energy.

\subsection{Differential apsidal precession}
\label{sec:GR}

As the debris passes through pericenter, it is deflected onto a self-intersecting trajectory by relativistic apsidal precession, leading to a stream-stream collision (Sec.~\ref{sec:si_cond}). At each orbital energy, the debris has a small spread in angular momentum, so different fluid elements have slightly different pericenters, resulting in differential apsidal precession (DAP) across the in-plane width of the stream \citep{Evans&Kochanek1989}. DAP produces a spread in argument of periapse \citep{Bonnerot&Lu2022},
\begin{equation}
	\Delta \delta \omega \simeq 3\pi \left( \frac{r_{\rm g}}{r_{\rm p}} \right)^2 \frac{\Delta y'_{\rm p}}{r_{\rm g}} \simeq 0.34^\circ \beta^2 Q_6^{1/3} \left( \frac{\Delta y'_{\rm p}}{3~R_\odot} \right)
\end{equation}

The resulting modification to the fluid velocities effectively produces a kick in the $\vu{y}'$-direction at pericenter with magnitude $v_{\rm orb} \sin(\Delta \delta \omega / 2)$, which can cause the debris stream to geometrically fan \citep{Evans&Kochanek1989, Gafton&Rosswog2019}. The $\vu{x}'$-velocities are also modified, but only to second order in $y'$ because $v_{x'}$ decreases on both sides of the central fluid element.

The in-plane stream width at self-intersection due to DAP is approximately $r_{\rm si} \cdot \Delta \delta \omega$, which gives $74~R_\odot$ for the fiducial stream slice, dominating the width $12~R_\odot$ from Newtonian evolution alone (Fig.~\ref{fig:si}). We test this prediction more directly by adding the DAP kick at pericenter by hand in the affine model. The in-plane width at self-intersection of the resulting post-nozzle stream closely matches our estimate.

These calculations show that broadening induced by DAP must be included to accurately determine the outgoing stream properties at self-intersection, even for only modestly relativistic TDEs. We leave a more careful treatment of this effect to future work. 

Surprisingly, this in-plane expansion only produces a small change in the vertical stream height compared to our previous analysis (Sec.~\ref{sec:second_orb}) because chemical processes make the gas pressure only weakly sensitive to changes in density, effectively decoupling the vertical and in-plane evolution. 

DAP also acts on the star during the initial disruption. However, the stream is initially confined by self-gravity, so rather than increasing the in-plane stream width, we expect it to modify the phase and amplitude of the quadrupolar oscillations.

\subsection{Nodal precession}
\label{sec:nodal}

If the BH has angular momentum, then the stream is deflected onto a different orbital plane by relativistic nodal precession, leading to a vertical offset at the self-intersection or even a complete miss \citep{Kochanek1994, Dai+2013, Guillochon&RamirezRuiz2015, Hayasaki+2016, Jankovic+2024}, delaying the self-intersection for several orbits \citep{Stone&Loeb2012}. 

Consider a BH with dimensionless spin $\chi$ and a debris stream with inclination $i$ relative to the BH spin. We estimate the per-orbit advance in the argument of periapse $\omega$ and the longitude of the ascending node $\Omega$ including de Sitter terms, Lense-Thirring terms, and quadrupole terms \citep{Merritt2013, Guillochon&RamirezRuiz2015}. The advance in the longitude of the ascending node rotates the angular momentum of the debris stream by $\delta \Omega \sin i$. At the self-intersection radius, the vertical offset of the streams is $\Delta z_{\rm si} \approx r_{\rm si} \delta \Omega \sin i$. 

For mildly relativistic disruptions $r_{\rm g} / r_{\rm p} \lesssim 0.02$, the vertical offset increases with $r_{\rm g} / r_{\rm p}$ because there is greater nodal precession. However, for more relativistic disruptions, the trend reverses because greater apsidal precession shrinks the self-intersection radius. 

\begin{figure*}
    \centering
    \includegraphics[width=0.67\linewidth]{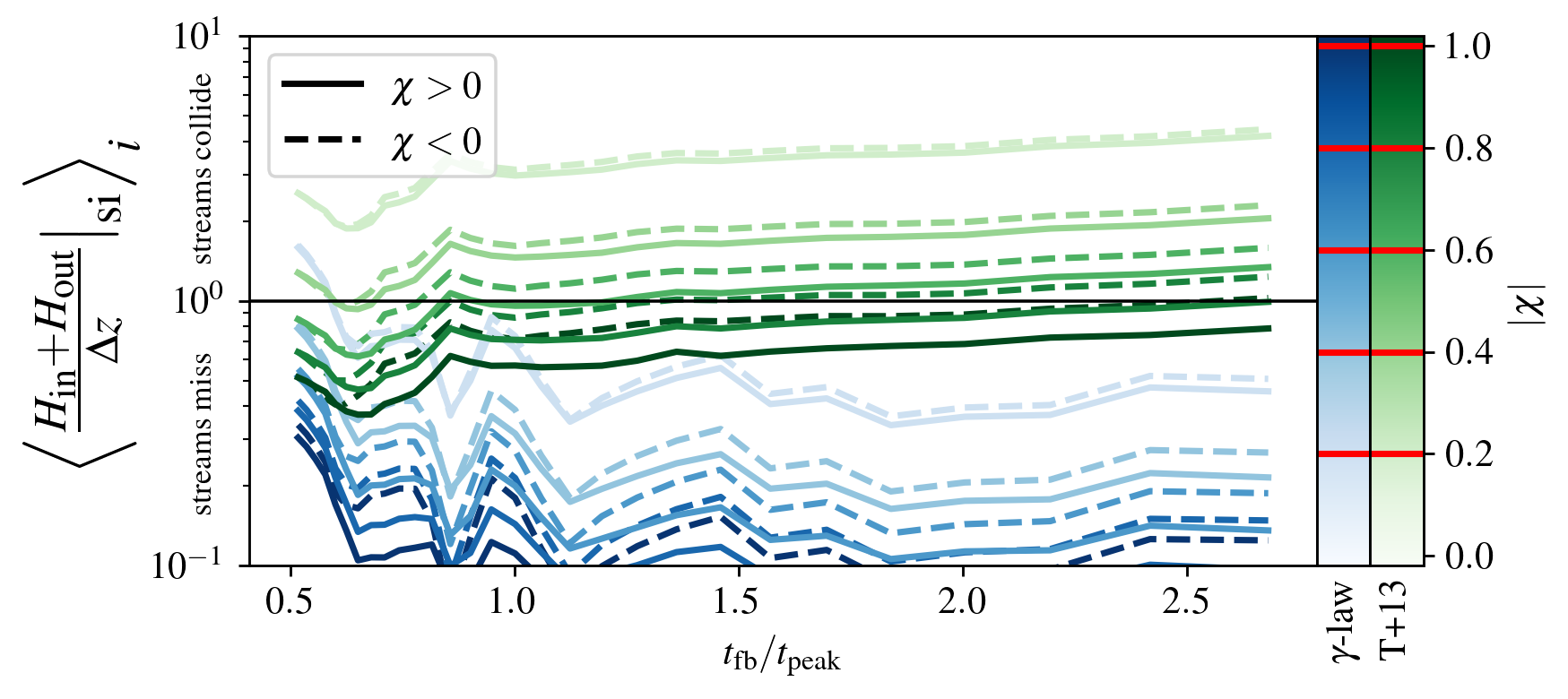}
    \caption{The ratio of the total vertical stream width to the vertical offset at self-intersection for several BH spins and all stream slices as a function of fallback time relative to peak. We average over an isotropic distribution of prograde (solid lines) and retrograde (dashed lines) inclinations. We show results with $\gamma$-law (blue lines) and \citetalias{Tomida+2013} (green lines) EOSs. We indicate the BH spin parameter by its color the color bar for the corresponding EOS. If this ratio is less than unity, then the stream will miss the first self-intersection for most inclinations. If this ratio is greater than unity, then its reciprocal indicates the typical relative vertical offset of the streams at self-intersection. With a $\gamma$-law EOS, complete misses are likely for all stream slices, even for small BH spins. With a \citetalias{Tomida+2013} EOS, misses are only likely for high BH spins $\gtrsim 0.6$ and $\lesssim < -0.8$. Even when the streams do not miss, there is likely to be a significant vertical offset. }
    \label{fig:nodal}
\end{figure*}

In Figure~\ref{fig:nodal}, we show the ratio of the total vertical stream width to the vertical offset at self-intersection for several BH spins and all stream slices as a function of fallback time relative to peak. We average over an isotropic distribution of prograde and retrograde inclinations. If this ratio is less than unity, then the stream will miss the first self-intersection for most inclinations. If this ratio is greater than unity, then its reciprocal indicates the typical relative vertical offset of the streams at self-intersection.

A complete miss is more likely for prograde orbits, which tend to enhance the nodal precession and reduce apsidal precession. With a $\gamma$-law EOS, complete misses are likely for all stream slices, even for small BH spins. With a \citetalias{Tomida+2013} EOS, misses are only likely for high BH spins $\gtrsim 0.6$ and $\lesssim -0.8$. Even when the streams do not miss, there is likely to be a significant vertical offset.

The chance of missing decreases at later fallback times because later stream slices spend more time in the hydrostatic regime on their first orbit, where they are inflated by continuous ${\rm H_2}$ formation (Fig.~\ref{fig:nodal}). For some carefully chosen BH spins, the self-intersection would only be missed at early times, so there could be a discontinuous change in the efficiency of circularization. 

There will also be differential nodal precession, which effectively rolls the debris stream about $\vu{x}'$ as different fluid elements are deflected onto different orbital planes \citep{Bonnerot&Lu2022}. The effect is only significant for more relativistic TDEs with high spins, because the leading-order term in the nodal precession angle scales $\propto (r_{\rm g} / r_{\rm p})^{3/2}$ compared to $\propto r_{\rm g} / r_{\rm p}$ for the apsidal precession angle. 

\subsection{Emission from the self-intersection point}
\label{sec:self_intersection}

Here we consider the simplified case of identical, cylindrical incoming and outgoing streams with zero vertical offset. The self-intersection shock converts the combined kinetic energy of the streams into radiation. If the optical depth at the shock were small, then the radiation would escape, producing a luminosity $L_0 \simeq 2 \dot{M}_{\rm fb} v_{r, {\rm si}}^2 \simeq 10^{43} \beta^3 Q_6^2~{\rm erg/s}$ (Eq.~\ref{eq:vrsi}).

In practice, the shock region is optically thick \citep{Lu&Bonnerot2020}, with an optical depth
\begin{equation}
    \tau_{\rm sh} \simeq \kappa_{\rm s} \rho_{\rm si} H_{\rm si} \simeq \frac{\kappa_{\rm s} \dot{M}_{\rm fb}}{\pi H_{\rm si} v_{r, {\rm si}}} \simeq 10^4\frac{\dot{M}_{\rm fb}}{M_\odot/{\rm yr}} \frac{30~R_\odot}{H_{\rm si}} \frac{0.01c}{v_{r, {\rm si}}}
\end{equation}
where $\rho_{\rm si}$ and $H_{\rm si}$ are the density and cylindrical radius of the stream at self-intersection. We have used typical values at self-intersection with a \citetalias{Tomida+2013} EOS (Fig.~\ref{fig:si}). The fallback rate is related to the stream properties by $\dot{M}_{\rm fb} \simeq \pi H_{\rm si}^2 \rho_{\rm si} v_{r, {\rm si}}$. The gas is fully ionized by the self-intersection shock, so we use the electron scattering opacity $\kappa_{\rm s} \simeq 0.34~{\rm cm^2/g}$. 

The high optical depth implies that the radiation is initially trapped. Internal energy is quickly converted into quasi-spherical bulk motion by adiabatic expansion, forming a wind with density profile $\rho(r) = 2 \dot{M}_{\rm fb} / (4\pi r^2 v_{\rm w})$, where $v_{\rm w} \sim v_{r, {\rm si}}$ is the wind velocity.

The radiation remains trapped until the effective photon diffusion velocity $c/\tau_{\rm w} \simeq c /\kappa \rho(r) r$ is comparable to the wind velocity, which occurs at the trapping radius \citep{Strubbe&Quataert2009}
\begin{equation}
    r_{\rm tr} \simeq \frac{\dot{M}_{\rm fb} \kappa_{\rm s}}{4\pi c} \simeq 800 \frac{\dot{M}_{\rm fb}}{M_\odot/{\rm yr}} R_\odot \simeq 16 \frac{\dot{M}_{\rm fb}}{M_\odot/{\rm yr}} r_{\rm p}
\end{equation}

At the trapping radius, the remaining internal energy is radiated away. The specific internal energy of radiation scales as $\varepsilon \propto \rho^{1/3} \propto r^{-2/3}$, so the resulting luminosity is
\begin{equation}\begin{split}
    L =\ & L_0 \left(\frac{r_{\rm tr}}{H_{\rm si}}\right)^{-2/3}\\
    \simeq\ & 10^{42}\left(\frac{\dot{M}_{\rm fb}}{M_\odot/{\rm yr}}\right)^{1/3} \left( \frac{H_{\rm si}}{30~R_\odot}\right)^{2/3} \beta^3 Q_6^2~{\rm erg/s}
    \label{eq:lum_si}
\end{split}\end{equation}

For our TDE parameters ($\beta = 2$, $Q_6 = 1$), the luminosity is slightly below observed TDE peak luminosities $10^{44}{\rm erg/s}$ \citep{Hinkle2020}. The increase in stream thickness by a factor $\sim 5$ due to chemical processes only mildly boosts the luminosity of the self-intersection shock, since $L \propto H_{\rm si}^{2/3}$.

For more massive SMBHs, such as the $10^7~M_\odot$ BH considered by \citet{Huang+2023}, or deeper disruptions, Equation~\ref{eq:lum_si} gives significantly higher luminosities due to the larger stream velocities. However, we have only considered the most optimistic case of identical streams with zero vertical offset. Our earlier results suggest that most self-intersections occur with significant differences in incoming and outgoing stream widths (Fig.~\ref{fig:si}) and vertical offsets (Fig.~\ref{fig:nodal}). Alternatively, the luminosity may have a significant, or even dominant, contribution from secondary circularization and accretion shocks in the outflow as suggested by \citet{Lu&Bonnerot2020}.

\subsection{Deviations from homologous evolution}
\label{sec:dev_homo}

The affine model is a first order expansion of the fluid equations about the center of each stream slice in the orbital plane. This is a good approximation for the tidal force as long as $H / r_{\rm c} \ll 1$. However, the approximation is less valid for the pressure gradient and self-gravitational forces in the outer layers of the stream. The affine model solutions should only be trusted at small cylindrical radii. Likewise, our \textsc{Athena++} simulations should only be trusted at small vertical heights. The largest errors occur near the surface of the stream, which affects the breakout of the outward-propagating shock launched at the nozzle.

For our implementation of arbitrary EOSs in the affine model, we assume that $K$ is a constant throughout the vertical stream slice profile. Ideally, $K$ should evolve independently for each fluid element according to the EOS. H recombination likely begins in the colder outer layers of the stream and gradually moves to smaller cylindrical radii as a recombination front \citep{Kasen&RamirezRuiz2010, Coughlin+2016b, Coughlin2023}. This creates a positive entropy gradient with cylindrical radius that reduces the negative pressure gradient in the debris stream. This effect could reduce the impact of H recombination on the stream widths, although the importance of the effect depends on how fast the recombination front reaches the center of the debris stream. 

Deviations from homologous evolution desynchronize the stream collapse at ballistic focal points. However, if the largest deviations occur in the outer layers of the stream, then the homologous description still captures the key dynamics at the nozzle \citep{Stone+2013}.

\subsection{Radiative cooling}
\label{sec:rad_cool}

As the debris stream expands, its optical depth decreases because the density falls faster than the stream width increases. H recombination further decreases the optical depth by removing free electrons from the gas, which dominate the opacity in ionized gas through scattering and bound-free transitions.

If the stream becomes optically thin, the recombination energy could radiate away, forcing the stream to shrink to maintain hydrostatic balance \citep{Coughlin+2016b}. \citet{Kasen&RamirezRuiz2010} hypothesized that the resulting emission would produce a recombination transient. However, \citet{Coughlin2023} argue that \citet{Kasen&RamirezRuiz2010} underestimate the optical depth because they neglect the hydrostatic phase of the debris stream evolution, and thus overestimate the stream expansion. Assuming an opacity from electron scattering, \citet{Coughlin2023} and \citet{Steinberg&Stone2024} estimate stream optical depths $\gtrsim 10^6$ at recombination.

Even if the stream does not become optically thin during recombination, it could still become optically thin at later times on the first orbit. In the ballistic regime, radiative cooling would not cause the stream to shrink, but the loss of internal energy would lead to more dissipation at the nozzle (Fig.~\ref{fig:series}).

We estimate the stream opacity using detailed low-temperature opacity tables from \textsc{Aesopus} \citep{Marigo&Aringer2009, Marigo+2022, Marigo+2024} with solar abundances \citep{Magg+2022}, metallicity $Z=0.02$, and Hydrogen mass fraction $X=0.7$ consistent with our \citetalias{Tomida+2013} EOS. Using the density and temperature evolution of the fiducial stream slice for the \citetalias{Tomida+2013} EOS, we bi-linearly interpolate the logarithmic Rosseland mean opacity $\log \kappa_{\rm R}$ in $\log T$-$\log R$ space, where $R \equiv \rho / T_6^3$ and $T_6 \equiv T / 10^6~{\rm K}$. 

After apocenter, the opacity is roughly constant at $\kappa_{\rm R} \simeq 0.040~{\rm cm^2/g}$, roughly an order-of-magnitude lower than the electron scattering opacity for ionized gas $0.34~{\rm cm^2/g}$. The dominant contributions to the opacity in the neutral stream are bound-free and free-free transitions in $H^-$ ions and absorption features in molecules like ${\rm H_2 O}$ and ${\rm TiO}$. These effects are not included in the opacity calculation of \citet{Kasen&RamirezRuiz2010}, leading them to underestimate the opacity by several orders of magnitude.

Using typical values near apocenter, the photon diffusion time is
\begin{equation}
	t_{\rm diff} \sim \frac{\rho \kappa_{\rm R} H}{c} H \simeq 1.6~\left(\frac{H}{30~R_\odot}\right)^2 \frac{\rho}{10^{-9}~{\rm g/cm^3}} \frac{\kappa_{\rm R}}{0.04~{\rm g/cm^2}}~{\rm hr}
\end{equation}
This is short compared to the orbital period of $26~{\rm days}$. However, on the first orbit, the stream is highly gas-pressure dominated $\beta_{\rm rad} \gtrsim 2\times 10^4$, so the radiative cooling timescale is significantly longer, $t_{\rm cool} \sim \beta_{\rm rad} t_{\rm diff} \gtrsim 3.7~{\rm yr}$. This exceeds the orbital period by more than an order-of-magnitude, so our adiabatic treatment is reasonable.

Due to dissipation at the nozzle, the debris stream is hotter $\sim 2\times 10^4~{\rm K}$ on the second orbit, and thus there is more energy in radiation, $\beta_{\rm rad} \gtrsim 200$. However, the opacity is also higher because the gas is ionized, so the cooling timescales are similar to the first orbit. The cooling timescale only becomes comparable to the orbital period if the stream misses the self-intersection and returns to pericenter on the second orbit. 

\subsection{Magnetic fields}
\label{sec:magnetic_fields}

Sun-like stars have typical surface magnetic field amplitudes from $1-10~{\rm G}$. There is no direct measurement of magnetic field amplitudes in stellar interiors. The strongest constraints for the Sun come from helioseismology, which give $\lesssim 300~{\rm kG}$, but these are only loose upper bounds \citep{Rashba+2007}. The initial field amplitude $B_0$ in the debris stream could fall anywhere in this range. The earliest fallback $\lesssim 0.4~t_{\rm peak}$ would sample the outer convective zone of a sun-like star $\gtrsim 0.7~R_\odot$ where magnetic fields are expected to be much weaker than the interior. 

3D magneto-hydrodynamic simulations show that the $\vu{x}'$-component of the magnetic field is amplified by tidal stretching due to magnetic flux conservation \citep{Bonnerot+2017, Guillochon&McCourt2017, Pacuraru+2025}, producing a field strength $B = B_0 R_*^2 / (H \Delta)$ \citep{Bonnerot&Lu2022}.

As the stream returns to pericenter and gets compressed, the magnetic pressure $B^2 / 8\pi$ grows as $\propto (H \Delta)^{-2}$, while gas and radiation pressure grow more slowly as $\propto (H \Delta)^{-\Gamma_1}$. Therefore, the impact of magnetic fields is maximized at the nozzle. For the fiducial stream slice, magnetic pressure exceeds gas and radiation pressure at the nozzle for an initial field amplitude,
\begin{equation}
    B_0 \gtrsim 150~\left( \frac{p_{\rm gas} + p_{\rm rad}}{10^{10}~{\rm erg/cm^3}} \right)^{1/2} \frac{H_{\rm noz}}{10^{-4}~R_\odot} \frac{\Delta_{\rm noz}}{3~R_\odot}~{\rm G}
\end{equation}
We have used the pressure just before the nozzle shock $10^{10}~{\rm erg/cm^3}$, which is smaller than the pressure at the nozzle shock by a factor $10^4$. This calculation suggests that magnetic pressure may reduce the vertical collapse at the nozzle, and thus the dissipation, relative to our results with no magnetic fields. Therefore, our results should be considered the limit of a weak initial field.

In a recent calculation, \citet{Pacuraru+2025} show that for strong $\gtrsim 10~{\rm kG}$ initial fields, magnetic pressure drives debris stream expansion during the first orbit. However, they acknowledge that this expansion may be overpowered by H recombination, which they do not include. 

They also show that H recombination may initiate Ohmic dissipation in the stream if the electron fraction falls below $x_{\rm e} = n_{\rm e} / n_{\rm n} \approx 10^{-14}$. 
Using the \citetalias{Tomida+2013} EOS, this condition is met by the fiducial stream slice when it recedes past $120~r_{\rm p}$ on the first orbit. However, the \citetalias{Tomida+2013} EOS does not include metals, which dominate the free electron population at the stream temperature around $3000~{\rm K}$ due to low ionization potential species like Na, C, and K.

We estimate the electron fraction using a simplified Saha model. For each species, we have,
\begin{equation}
    \frac{n_{i+}}{n_{i0}} = \frac{2 }{n_{i, {\rm e}}} \frac{e^{-\chi_i / k_{\rm B} T}}{\lambda_{\rm th}^3}
\end{equation}
where $n_{i+}$ and $n_{i0}$ are the number densities in the ionized and neutral states respectively, $\chi_i$ is the ionization energy, and $\lambda_{\rm th}^2 = h^2/(2\pi m_{\rm e} k_{\rm B} T)$ is the electron thermal de Broglie wavelength. Summing the free electron contributions, we find,
\begin{equation}
    x_{\rm e}^2 = \frac{1}{\lambda_{\rm th}^3} \frac{1}{n_{\rm H}} \sum x_i e^{-\chi_i / k_{\rm B} T}
\end{equation}
Using $x_{\rm Na} \sim 2\times 10^{-6}$, $x_{\rm C} \sim 3\times 10^{-4}$, $x_{\rm K} \sim 10^{-7}$, $\chi_{\rm Na} \simeq 5.14~{\rm eV}$, $\chi_{\rm C} \simeq 11.3~{\rm eV}$, $\chi_{\rm K} \simeq 4.34~{\rm eV}$, and $T=3000~{\rm K}$ gives
\begin{equation}
    x_{\rm e} \simeq 10^{-4}\left( \frac{\rho}{10^{-9}~{\rm g/cm^3}}\right)^{-1/2}
\end{equation}
which is consistent with the other estimates in the literature \citep{Gammie&Menou1998}. These results suggest that Ohmic resistivity never becomes important in the debris stream.

%% file: sections/05_conclusion.tex
The emission from TDEs is powered by a combination of circularization and accretion of bound stellar debris.  In this paper we analyze the role of the nozzle shock produced as the debris returns to pericenter and undergoes extreme vertical compression. We use an idealized model combining 3D SPH simulations and the semi-analytic affine model to generate initial conditions for 1D finite-volume hydrodynamic simulations which can unambiguously resolve the nozzle shock. Because our model is computationally cheap, we can analyze the debris stream dynamics across a wide range of fallback times using a realistic EOS. We analyze the role of H recombination and ${\rm H_2}$ formation in expanding the stream prior to the nozzle shock, the formation of the nozzle shock and its associated energy dissipation, and the interplay between pressure gradient and tidal forces that set the properties of the stream at the self-intersection point. We summarize our findings below.

\begin{enumerate}[label=(\roman*)]

\item
\textit{Pre-nozzle evolution and the role of recombination}: The stream width after the disruption is initially set by a balance between pressure gradient and self-gravitational forces (hydrostatic regime), but eventually tidal forces dominate (ballistic regime) (Fig.~\ref{fig:evol}). The ballistic trajectories of all fluid elements intersect at vertical and in-plane focal points, compressing the stream against pressure gradient forces. The vertical focal point near pericenter, where tides are the strongest, is responsible for the nozzle shock. H recombination broadens the stream by a factor $\sim 5$ and ${\rm H_2}$ formation adds an additional order-unity factor (Fig.~\ref{fig:dim}). This increases the energy available to dissipate at the nozzle and maintains a stream temperature around $3000~{\rm K}$ (Fig.~\ref{fig:phasespace}). The most bound debris enters the ballistic regime at earlier times, so it is less affected by the increase in pressure due to H recombination. Chemical processes should be included in future global simulations of TDEs.

\item
\textit{The nozzle shock}: At the nozzle, the debris stream reaches a vertical width $10^{-4}~R_\odot$ (Fig.~\ref{fig:profile}) before it is reversed by pressure gradients, with similar contributions from gas and radiation. We do not include magnetic pressure, whose contribution depends on the amplitude of the initial stellar field, which is highly uncertain (Sec.~\ref{sec:magnetic_fields}). Before maximum compression, a shock forms and propagates towards the center of the stream, where it collides with the shock from the other side of the $xy$-plane (Fig.~\ref{fig:lag}). The collision launches outward-propagating shocks that reverse the collapse until they break out of the stream surface. The gas is ionized by the inward-propagating shock and all fluid elements have a similar net entropy increase after both shock encounters. The entire interaction takes place over seconds, a small fraction of the orbital period of weeks to months. To accurately resolve the nozzle shock in a Lagrangian code, the required mass resolution is $\lesssim 10^{-9}~M_\odot$ at peak fallback for our TDE parameters (Sec.~\ref{sec:3d_imp}).

\item
\textit{Post-nozzle evolution}: The orbital planes of all fluid elements intersect along a line. The fluid elements are ballistically focused towards the $xy$-plane where the intersection line crosses the orbit: once at the nozzle and once at larger radii. At the second focal point, pressure gradients, enhanced by dissipation at the nozzle, resist the collapse and drive significant stream expansion (Fig.~\ref{fig:orbits}). This makes the post-nozzle stream widths sensitive to pressure, and therefore to dissipation at the nozzle. With a realistic EOS, the second focal point occurs close to the self-intersection radius, so this expansion affects the outgoing stream properties at self-intersection. This result depends sensitively on the dynamics during the first orbit, so the post-nozzle stream evolution may need to be analyzed for each new TDE setup on a case-by-case basis.

\item
\textit{Geometry of the self-intersection}: 
The expansion of the stream due to H recombination and nozzle shock dissipation increases the likelihood that the self-intersection occurs, despite relativistic nodal precession. In our simulations, the stream is only likely to miss the self-intersection for dimensionless BH spins $\gtrsim 0.6$ and $\lesssim -0.8$ (Fig.~\ref{fig:nodal}). However, for most BH spins and inclinations, there is still a significant vertical offset. We also find that at self-intersection, the outgoing stream is thinner than the incoming stream by a factor of a few, suggesting that only a fraction of the stream participates in the self-intersection (Fig.~\ref{fig:si}). However, we caution that the in-plane stream width of the outgoing stream may be significantly enhanced by differential apsidal precession, which we do not include here (Sec.~\ref{sec:GR}). 

\item
\textit{Emission from the self-intersection}: 
The luminosity generated at the self-intersection can explain observed TDE luminosities for our TDE parameters ($\beta=2$, $Q=10^6$), but only for the optimistic case of identical streams and zero vertical offset. 
The expansion of the stream due to chemical processes may increase the luminosity by lowering the optical depth of the self-intersection outflow (Eq.~\ref{eq:lum_si}), but only by a factor of a few.
Alternatively, the luminosity may have a significant, or even dominant, contribution from secondary circularization and accretion shocks in the outflow launched at self-intersection.

\end{enumerate}

Our results clarify the nozzle shock's role in circularization in TDEs, providing a foundation for more realistic circularization and emission models. The main limitation of our work is that our results may be affected by differential apsidal precession (Sec.~\ref{sec:GR}) and magnetic fields (Sec.~\ref{sec:magnetic_fields}), which we do not include. These effects, and their interplay with chemical process and ballistic dynamics, should be carefully analyzed by future work.

Other parts of the TDE parameter space should also be explored. Disruptions by more massive SMBHs up to $10^7~M_\odot$ produce a similar volumetric TDE rate to the $10^6~M_\odot$ case \citep{Yao+2023}. Although more deeply penetrating TDEs are more rare, some will be observed by LSST and may exhibit unique signatures due to more extreme compression at the nozzle. 

%% file: sections/06_acknowledgements.tex
Z.A. would like to acknowledge helpful conversations with Cl{\'e}ment Bonnerot, Shane Davis, Xiaoshan Huang, Yan-Fei Jiang, Wenbin Lu, Valeriia Rohoza, Elad Steinberg, Jim Stone, and Romain Teyssier.

This work benefited from interactions supported by the Gordon and Betty Moore Foundation through grant GBMF5076.  This research was supported in part by grant NSF PHY-2309135 to the Kavli Institute for Theoretical Physics (KITP). This material is based upon work supported by the U.S. Department of Energy, Office of Science, Office of Advanced Scientific Computing Research, Department of Energy Computational Science Graduate Fellowship under Award Number DE-SC0024386.

This report was prepared as an account of work sponsored by an agency of the United States Government. Neither the United States Government nor any agency thereof, nor any of their employees, makes any warranty, express or implied, or assumes any legal liability or responsibility for the accuracy, completeness, or usefulness of any information, apparatus, product, or process disclosed, or represents that its use would not infringe privately owned rights. Reference herein to any specific commercial product, process, or service by trade name, trademark, manufacturer, or otherwise does not necessarily constitute or imply its endorsement, recommendation, or favoring by the United States Government or any agency thereof. The views and opinions of authors expressed herein do not necessarily state or reflect those of the United States Government or any agency thereof.

The simulations presented in this article were performed on computational resources managed and supported by Princeton Research Computing, a consortium of groups including the Princeton Institute for Computational Science and Engineering (PICSciE) and the Office of Information Technology’s High Performance Computing Center and Visualization Laboratory at Princeton University.

ERC acknowledges support from the National Aeronautics and Space Administration NASA through the Astrophysics Theory Program, grant 80NSSC24K0897. CJN acknowledges support from the Leverhulme Trust (grant No.\ RPG-2021-380).

%% file: sections/07_data.tex
The data presented in this article will be made available upon
reasonable request to the corresponding author. The code for generating our \citetalias{Tomida+2013} EOS is available online (Footnote~1).

%% file: sections/08_appendix.tex
\section{Toy model for debris stream evolution}
\label{sec:toy_model}

In this appendix, we summarize a simplified model for the evolution of the debris stream developed across several previous works \citep{Kochanek1994, Coughlin+2016a, Coughlin+2016b, Bonnerot&Lu2022, Bonnerot+2022, Coughlin2023}. We present the relevant details here both for completeness and because they provide useful context for interpreting our results. 

After the initial disruption, the tidal field elongates the stellar debris into a quasi-cylindrical debris stream. Along the axis of the cylinder, tidal forces dominate over self-gravitational and pressure gradient forces, so the stream is well-described as a set of independently evolving cross sections, or stream slices, moving on Keplerian orbits with high eccentricities $1-e \simeq 10^{-2}$. In the transverse directions, the self-gravitational and pressure gradients forces are initially large enough that the stream is in quasi-hydrostatic equilibrium \citep{Coughlin+2016a, Coughlin+2016b}. 

The stream slice expands to maintain equilibrium as the density decreases due to longitudinal stretching. Eventually, tidal forces dominate in the transverse directions, and the subsequent evolution is ballistic for the remainder of the orbit. We refer to these phases of the evolution as the hydrostatic and ballistic regimes. 

Due to its high eccentricity, the stream slice orbit can be approximated as radial, for which $r = (9G M_\bullet t^2 / 2)^{1/3}$. The stream thickness evolves due to the tidal force according to $\partial_t^2 L \propto 2 G M_\bullet L / r^3$. These equations admit two solutions $L \propto r^2$ and $L \propto r^{-1/2}$ which describe the evolution before and after apocenter respectively \citep{Sari+2010, Bonnerot+2022}.

The tidal, pressure gradient, and self-gravitational forces in the stream slice plane are approximately,
\begin{align}
	a_{\rm t} =\ & (GM_\bullet/r^3) S \propto S / r^3\\
	a_{\rm p} =\ & (1/\rho) \partial_s p \sim (1 / \rho) p / S \propto \rho^{\Gamma_1 - 1} / S\\
	a_{\rm g} =\ & \partial_s \phi \sim 2\pi G S \rho \propto S \rho
\end{align}
where for the moment we assume $H = \Delta \equiv S$. 

In the hydrostatic regime, the pressure gradient and self-gravitational forces are balanced $a_{\rm p} \sim a_{\rm g}$, which implies $S^2 \propto \rho^{\Gamma_1 - 2}$. Combining this with the scalings for $\rho$ and $L$ before apocenter, we find $\rho \propto r^{-2/(\Gamma_1 - 1)}$ and $S \propto r^{(2 - \Gamma_1) / (\Gamma_1 - 1)}$ \citep{Coughlin+2016b, Coughlin2023}. For $\Gamma_1 = 5/3$, this gives $\rho \propto r^{-3}$ and $S \propto r^{1/2}$, indicating that the stream slice expands as it moves towards apocenter. As $\Gamma_1$ approaches unity, the expansion accelerates because the pressure gradient forces become insensitive to changes in density. Repeating the same calculation after apocenter, we find $\rho \propto r^{1/[2(\Gamma_1 - 1)]}$ and $S \propto r^{(2 - \Gamma_1) / [4(\Gamma_1 - 1)]}$. For $\Gamma_1 = 5/3$, this gives $\rho \propto r^{3/4}$ and $S \propto r^{-1/8}$, indicating that the stream slice continues to expand as it falls back towards pericenter, albeit gradually.

The stream slice enters the ballistic regime when $a_{\rm t} \gtrsim a_{\rm p}, a_{\rm g}$, which occurs at a critical density $\rho_\bullet = M_\bullet / (2\pi r^3)$, which we refer to as the BH density following \citet{Coughlin+2016b}. For $\Gamma_1 = 5/3$, the density and the BH density have the same scaling before apocenter, suggesting that the transition to the ballistic regime is delayed until apocenter. In the hydrostatic regime, the stream slice always expands, so $\partial_t S$ is positive at the transition to the ballistic regime. In the ballistic regime, the stream width evolves due to the tidal force according to $\partial_t^2 S \propto -G M_\bullet S / r^3$. The tidal force is always negative, so eventually $\partial_t S$ changes sign. We refer to this point as the equilibrium point \citep{Bonnerot+2022}. 

Similar to the stream thickness, the equations for $r(t)$ and $\partial_t^2 S$ admit two solutions $S\propto r^{1/2}$ and $S\propto r$ \citep{Sari+2010, Bonnerot+2022}. \citet{Bonnerot&Lu2022} show that when the vertical width follows the $r^{1/2}$ (resp. $r$) solution when the stream slice moves along its orbit parallel (resp. perpendicular) to the intersection line. Similarly, the in-plane width can take either scaling depending on the orbital motion of the stream slice relative to the in-plane focal point. We can estimate a range of possible scaling relations by considering the set of scalings for $H$ and $\Delta$. Before apocenter, the density scales between $\rho\propto r^{-3}$ and $\propto r^{-4}$. After apocenter, the density scales between $\rho \propto r^{-1/2}$ and $\propto r^{-3/2}$.

Before apocenter, the ratio of pressure gradient to tidal forces scales between $a_{\rm p} / a_{\rm t} \propto r^{5 - 3 \Gamma_1}$ and $\propto r^{5 - 4 \Gamma_1}$. The sign of the exponent depends on the value of $\Gamma_1$. Pressure gradient forces become stronger relative to tidal forces as the stream moves towards apocenter for $\Gamma_1 < 1.2$. Therefore, in this case pressure gradient forces may be comparable to tidal forces even after the stream slice enters the ballistic regime. Repeating the same calculation after apocenter, we find $a_{\rm p} / a_{\rm t} \propto r^{(5 - \Gamma_1)/2}$ and $\propto r^{(5 - 3 \Gamma_1)/2}$. The sign of the exponent is never negative for $\Gamma_1 \le 5/3$, so tidal forces always remain dominant. 

\section{Affine model equations}
\label{sec:affine_eqs}

In this appendix, we describe our version of the debris stream affine model developed by \citet{Kochanek1994} and refined by \citet{Bonnerot+2022}. In Section~\ref{sec:affine}, we highlighted how our version of the model differs from these previous works. Here, we present the full model equations.

Let $\vb{r}_{\rm c}(\varepsilon)$ be the center-of-mass position of the stream slice with specific orbital energy $\varepsilon$. We define a vector $\vell \equiv -\partial \vb{r}_{\rm c} / \partial \varepsilon$ tangential to the line joining the stream slice centers of mass. Using this vector, we construct a convenient non-inertial coordinate system $(x', y', z')$ with origin $\vb{r}_{\rm c}$ and stream slice orthogonal to $\vu{x}'$. Specifically, the unit vectors are given by $\vu{x}' \equiv \vell / \ell$ , $\vu{y}' \equiv \vu{z} \cross \vu{x}'$, and $\vu{z}' = \vu{z}$ where $\ell \equiv \norm{\vell}$ (Fig.~\ref{fig:diagram}).

It is convenient to define a vector $\boldsymbol{\Gamma} \equiv \partial_t \vell / \ell$, which can be decomposed as $\boldsymbol{\Gamma} = \lambda \vu{x}' + \Omega \vu{y}'$, where $\lambda \equiv \partial_t \ell / \ell = \partial_t \vell \vdot \vu{x}' / \ell$ and $\Omega \equiv \partial_t \vell \vdot \vu{y}' / \ell$ denote the rate of longitudinal stretching and the angular frequency of stream slice rotation respectively. The angle $\alpha$ between $\vu{x}'$ and $\vu{x}$ evolves according to $\partial_t \alpha = \Omega$. The center-of-mass of each stream slice moves on a Keplerian orbit which can be followed analytically.

Due to the tidal force, $\vell$ evolves according to \citep{Kochanek1994}
\begin{equation}
    \pdv[2]{\vell}{t} = -\frac{G M_\bullet}{r_{\rm c}^3} \left( \vell - 3 \frac{\vb{r}_{\rm c} \vdot \vell}{r_{\rm c}^2} \vb{r}_{\rm c} \right)
    \label{eq:ell_ddot}
\end{equation}
Let $\theta$ denote the angle between $\vu{r}_{\rm c}$ and $\vu{x}$ and let $\delta \equiv \alpha - \theta$ denote the angle between $\vu{r}_{\rm c}$ and $\vu{x}'$. In terms of $\boldsymbol{\Gamma}$, Equation~\ref{eq:ell_ddot} becomes
\begin{equation}
    \pdv{\boldsymbol{\Gamma}}{t} = -\frac{G M_\bullet}{r_{\rm c}^3} (\vu{x}' - 3 \cos \delta \vu{r}_{\rm c}) - \lambda \boldsymbol{\Gamma}
\end{equation}
Projecting onto the $\vu{x}'$ and $\vu{y}'$-directions, we find \citep{Bonnerot+2022}
\begin{align}
    \pdv{\lambda}{t} =\ & \Omega^2 - \lambda^2 - \frac{G M_\bullet}{r_{\rm c}^3} (1 - 3 \cos^2 \delta) \label{eq:lambda}\\
    \pdv{\Omega}{t} =\ & -2 \lambda \Omega + 3 \frac{G M_\bullet}{r_{\rm c}^3} \cos \delta \sin \delta \label{eq:Omega}
\end{align}

Let $H$, $\Delta$, and $L$ denote the height, width, and thickness of the stream slice in the $\vu{z}$, $\vu{y}'$, and $\vu{x}'$-directions respectively (Fig.~\ref{fig:diagram}). Let $\hat{H}$, $\hat{\Delta}$, and $\hat{L}$ denote the same quantities normalized by their initial values. In the affine model, the stream slice evolution is described by a linear transformation in the orbital frame. By construction, the transformation matrix is diagonal in our coordinate system with $\mathsf{M}={\rm diag}(\hat{L}, \hat{\Delta}, \hat{H})$. The density profile of the stream slice is scaled from the initial density profile as $\rho = \rho_0 /  {\rm det} (\mathsf{M}) = \rho_0 / (\hat{L} \hat{\Delta} \hat{H})$.

We define the entropy proxy $K \equiv p / \rho^\gamma$. Let $\hat{K}$ denote the entropy proxy normalized by its initial value. For a general EOS, $\hat{K}$ may evolve because the specific entropy is not necessarily proportional to $\ln K$. We assume that $K$ is constant throughout the stream slice profile. Ideally, $K$ should evolve independently for each fluid element according to the EOS, but this behavior cannot be captured in an affine model because it breaks the assumption of homologous evolution (Sec.~\ref{sec:dev_homo}). The entropy proxy evolves according to
\begin{equation}
	\pdv{K}{t} = (\Gamma_1 - \gamma) \frac{K}{\rho} \pdv{\rho}{t}
	\label{eq:Kam}
\end{equation}
where $\partial_t \rho = -\rho (\lambda + \partial_t \ln \hat{\Delta} + \partial_t \ln \hat{H})$ by the chain rule. We evolve the entropy proxy, rather than the specific internal energy \citep{Kochanek1994}, because the entropy is explicitly conserved in the limit $\Gamma_1 \to \gamma$. The temperature can be calculated from the density and entropy proxy using the EOS, or it can be evolved according to
\begin{equation}
	\pdv{T}{t} = (\Gamma_3 - 1) \frac{T}{\rho} \pdv{\rho}{t}
	\label{eq:tempam}
\end{equation}

The stream slice is subject to forces from pressure gradients, self-gravity, the tidal field, and pseudo-forces produced by our rotating coordinate system. Expanding to second order about the center of the stream slice, the pressure gradient forces are \citep{Coughlin2023}
\begin{equation}\begin{split}
    \frac{1}{\rho} \pdv{p}{y'} \simeq\ & \frac{\gamma \hat{K}}{\hat{\Delta} \rho_{\rm 0, c}} \eval{\pdv[2]{\rho_0}{(y')}}_{\rm c} \frac{p_{\rm c}}{\rho_{\rm c}} y'_0\\
    \frac{1}{\rho} \pdv{p}{z'} \simeq\ & \frac{\gamma \hat{K}}{\hat{H} \rho_{\rm 0, c}} \eval{\pdv[2]{\rho_0}{(z')}}_{\rm c} \frac{p_{\rm c}}{\rho_{\rm c}} z'_0
    \label{eq:pres_grad}
\end{split}\end{equation}
where subscripts 0 and c denote quantities evaluated at the initial conditions and the center of the stream slice respectively. 

If pressure and self-gravity are initially balanced in the stream slice, then the density profile can be described by a cylindrical polytrope with Emden index $n = 1/(\gamma - 1)$ satisfying the cylindrical Lane-Emden equation \citep{Ostriker1964},
\begin{equation}
    \frac{1}{\xi} \dv{\xi} \left( \xi \dv{\vartheta}{\xi} \right) = -\vartheta^n
    \label{eq:lane_emden}
\end{equation}
where $s = \alpha_{\rm le} \xi$ is the cylindrical radius, $\rho = \rho_{\rm c} \vartheta^n$ is the density, and $\alpha^2_{\rm le} \equiv (n+1)K / (4\pi G \rho_{\rm c}^{1 - 1/n})$. The boundary of the stream slice occurs when the density becomes negative, which occurs at $\xi_{\rm max} \simeq 2.6477$ for Emden index $n=1.5$.

The central second derivatives for a cylindrical polytrope profile are
\begin{equation}
    \eval{\pdv[2]{\rho_0}{(y')}}_{\rm c} = \eval{\pdv[2]{\rho_0}{(z')}}_{\rm c} = -\frac{n}{2} \frac{\rho_{\rm c, 0}}{\alpha^2} = -\frac{2\pi G}{\gamma} \frac{\rho_{\rm c, 0}^3}{p_{\rm c, 0}}
\end{equation}
Substituting these expressions into Equation~\ref{eq:pres_grad} yields
\begin{equation}
    \frac{1}{\rho} \pdv{p}{y'} \simeq -\frac{2\pi G \rho_{\rm c, 0} \hat{K}}{(\hat{L} \hat{H})^{\gamma-1} \hat{\Delta}^\gamma} y'_0, \quad \frac{1}{\rho} \pdv{p}{z'} \simeq -\frac{2\pi G \rho_{\rm c, 0} \hat{K}}{(\hat{L} \hat{\Delta})^{\gamma - 1} \hat{H}^\gamma} z'_0
\end{equation}

Making use of Gauss's theorem and approximating the circumference of the stream slice as $\pi (H + \Delta)$, the self-gravitational forces are \citep{Kochanek1994}
\begin{equation}
    \dv{\phi}{y'} \simeq 4\pi G \frac{\rho_{\rm c, 0}}{\hat{L} (\hat{H} + \hat{\Delta})} y_0', \quad \dv{\phi}{z'} \simeq 4\pi G \frac{\rho_{\rm c, 0}}{\hat{L} (\hat{H} + \hat{\Delta})} z'_0
\end{equation}
By definition, $\hat{L} = \hat{\Delta} = \hat{H} = \hat{K} = 1$ in the initial conditions. In this case, the forces from pressure and self-gravity are equal and opposite, consistent with our assumption of hydrostatic equilibrium in the initial conditions.

In the $\vu{z}$-direction, the equation of motion is
\begin{equation}
    \pdv[2]{z'}{t} = -\frac{1}{\rho} \pdv{p}{z'} - \pdv{\phi}{z'} - \frac{G M_\bullet}{r_{\rm c}^3} z' \label{eq:z}
\end{equation}
Substituting our expressions for the pressure gradient and self-gravitational forces and dividing by $z_0'$, we find
\begin{equation}
    \pdv[2]{\hat{H}}{t} = 2\pi G \rho_{\rm c, 0} \left[ \frac{\hat{K}}{(\hat{L} \hat{\Delta})^{\gamma - 1} \hat{H}^\gamma} - \frac{2}{\hat{L} (\hat{\Delta} + \hat{H})} \right] - \frac{G M_\bullet}{r_{\rm c}^3} \hat{H}
    \label{eq:H}
\end{equation}

In the $\vu{y}'$-direction, the equation of motion is
\begin{equation}
    \pdv[2]{y'}{t} = -\frac{1}{\rho} \pdv{p}{y'} - \pdv{\phi}{y'} - \frac{G M_\bullet}{r_{\rm c}^3} y' (1 - 3 \sin^2 \delta) - \Omega^2 y' - 2 \Omega v_\parallel
    \label{eq:yp_eom}
\end{equation}
where $v_\parallel$ is the velocity perpendicular to the stream slice plane. The last two terms are contributions from the pseudo-force due to the rotation of the coordinate system $\Omega^2 y'$ and the in-plane shear $-2\Omega (\Omega y' + v_\parallel)$ i.e. gas at $y' \ne 0$ which moves along the $\vu{x}'$-direction with respect to the center of the stream slice. The evolution of $v_\parallel$ is given by
\begin{equation}
    \pdv{v_\parallel}{t} = 3 \frac{G M_\bullet}{r_{\rm c}^3} y' \cos \delta \sin \delta + \pdv{y'}{t} \Omega - \lambda (\Omega y' + v_\parallel) \label{eq:vpar2}
\end{equation}
Substituting our expressions for the pressure gradient and self-gravitational forces and dividing by $y_0'$, we find
\begin{equation}\begin{split}
    \pdv[2]{\hat{\Delta}}{t} =\ & 2\pi G \rho_{\rm c, 0} \left[ \frac{\hat{K}}{(\hat{L} \hat{H})^{\gamma - 1} \hat{\Delta}^{\gamma}} - \frac{2}{\hat{L} (\hat{\Delta} + \hat{H})} \right]\\
    \ & - \frac{G M_\bullet}{r_{\rm c}^3} \hat{\Delta} (1 - 3 \sin^2 \delta) -\Omega^2 \hat{\Delta} - 2 \Omega \hat{v}_\parallel
    \label{eq:Delta}
\end{split}\end{equation}
\begin{equation}
    \pdv{{\hat{v}}_\parallel}{t} = 3 \frac{G M_\bullet}{r_{\rm c}^3} \hat{\Delta} \cos \delta \sin \delta + \pdv{\hat{\Delta}}{t} \Omega - \lambda (\Omega \hat{\Delta} + \hat{v}_\parallel) \label{eq:vpar}
\end{equation}
where $\hat{v}_\parallel$ denotes the parallel velocity normalized by its initial value. For a more detailed derivation and explanation of these equations, see Appendix~A of \citet{Bonnerot&Lu2022}.